# Evidence for a giant companion orbiting Tabby's star

Cristina Madurga-Favieres ,[1★] P. A. Strøm ,[1] Thomas G. Wilson ,[1] Neda Heidari [2] and Flavien Kiefer [3]

[1]*Department of Physics, University of Warwick, Gibbet Hill Road, Coventry CV4 7AL, UK*
[2]*Institut d'astrophysique de Paris, UMR 7095 CNRS université Pierre et Marie Curie, 98 bis, boulevard Arago, F-75014 Paris, France*
[3]*LESIA, Observatoire de Paris, Université PSL, CNRS, Sorbonne Université, Université de Paris, 5 place Jules Janssen, F-92195 Meudon, France*

## ABSTRACT

KIC 8462852, also known as Tabby's star, shows deep, irregular flux variations up to 20 per cent in its *Kepler* light curve. These unusual events show no apparent periodicity. Multiwavelength observations show the dips are consistent with transiting dust, although its origin remains uncertain. The leading explanation is that dust may originate from a family of exocomets or planetesimal fragments. *TESS* data reveal a unique, symmetric transit event consistent with a planet or brown dwarf. This would support the theory that incoming exocometary bodies may result from orbital perturbations by a planet. The analysis of the *TESS* and, archival and newly obtained radial velocity (RV) data, suggests that a companion could orbit around Tabby's star with a period of at least ∼1030 d. When combined with the constraints imposed by *Kepler* and other photometric surveys, we find only five orbital period windows are possible. We conduct observations to follow-up the first two windows, significantly reducing the allowed ranges. Fitting the *TESS* transit using exocometary models produces a poor fit or the need of physically unrealistic parameters, favouring the giant companion hypothesis. A 2.3$\sigma$ detection of $M = 9.4^{+4.9}_{-4.1}$ $M_{Jup}$ (3$\sigma$ upper limit $< 28.0$ $M_{Jup}$) at the transit orbital period provides tentative evidence of the companion, but more RV observations are required. Along with the radius $R = 1.7 \pm 0.1$ $R_{Jup}$, the companion would lie in the giant planet-brown dwarf regime. In the future *Gaia* DR4 astrometry will reveal the nature of this object.

## 1 INTRODUCTION

KIC 8462852, also known as Tabby's star, is a peculiar F3V main sequence star. Its unusual *Kepler* light curve has aroused interest over the last decade due to the numerous irregular and asymmetric dips it shows. T. S. Boyajian et al. (2016) found that the *Kepler* data from 2009 to 2013 revealed variations in brightness which included two main events: one isolated 15 per cent flux drop around day 800 (D800), that lasted almost a week, and one complex sequence of irregular photometric dips lasting 80 d, almost two years later (around day 1500, D1500), with the deepest absorption within the sequence reaching a depth of ∼22 per cent. These series of dips show no apparent periodicity, have different durations (by factor of three), depths (for example ∼0.5, ∼3, ∼8 per cent) and asymmetric shapes (T. S. Boyajian et al. 2016). For the remainder of the *Kepler* light curve, the star is relatively constant, with a few smaller ∼0.5 per cent dips earlier in the mission, making the dimming events even more peculiar. It was also found that the star dimmed at a rate of 0.3 per cent per year during the first three years of data, and later decreased its flux by 2.5 per cent over six months. It is the largest decrease found within the ∼550 nearby comparison stars and stars with similar stellar parameters identified in *Kepler* images by B. T. Montet & J. D. Simon (2016).

Since its discovery, efforts have been directed into revealing the source of these events with follow-up observations made at multiple wavelengths. *WISE* (T. S. Boyajian et al. 2016) and *Spitzer*/IRAC (M. Marengo, A. Hulsebus & S. Willis 2015) observations have independently shown the lack of a mid-infrared excess in the system. These results argue against an accretion disc, a gas-rich protoplanetary disc, or collisions of planetesimals in an asteroid belt as the sources of the intrinsic flux variability. Moreover, the known presence of a much fainter nearby (∼1.95 arcsec) M-dwarf star (*Gaia* DR3 2081900944807842560) (M2V), with a mass of $M = 0.44{\pm}0.02$ $M_{\odot}$ (L. A. Pearce et al. 2021), located within the same *Kepler* pixel, can only account for irregular dips up to ∼3 per cent. Additionally, it is highly unlikely that the brightness variations are caused by a large object colliding with Tabby's star, as the probability of such an event is extremely low (T. S. Boyajian et al. 2016). Observations taken almost simultaneously across multiple filters (central wavelengths of the bands at 556, 618, 709, 792, and 868 nm) have shown a wavelength dependent absorption depth (H. J. Deeg et al. 2018), suggesting that the dips are most likely caused by isolated collections of dust.

Among the proposed scenarios that would explain the irregularity of the wavelength dependent dips, the most plausible one is the presence of a family of exocomets or planetesimal fragments that break up close to the star and travel in highly eccentric orbits (T. S. Boyajian et al. 2016). When passing close to the peri-centre, suggested by the non-repetition of equal events (T. S. Boyajian et al. 2016), the dusty material causes photometric dimming (A. Lecavelier Des Etangs, A. Vidal-Madjar & R. Ferlet 1999). Even though individual comets may not produce the deepest dips seen in the *Kepler* light curve, a family of comets could produce a resulting transits that would be longer and deeper, similar to the observed phenomena in the *Kepler* data of Tabby's star (H. Beust et al. 1996). Moreover, the fragments of the family of comets would share very similar orbits, making it possible to detect multiple transits if the orbits are aligned in such a way that it favours transit detection. Thus, if this scenario is correct, the presence of one transit would strongly hint at the detectability of transits from other co-orbiting bodies. A model of circumstellar material distributed along an elliptical orbit can account for the absence of infrared excess at the current detection thresholds in the infrared light curves, and place the family of exocomets between 0.005 and 0.6 au (M. C. Wyatt et al. 2018). E. H. L. Bodman & A. Quillen (2016) fitted the *Kepler* light curve with dense clumps of numerous comets (∼100) that may explain some of the complex series of dips observed in Quarters 16 and 17 of the *Kepler* data. They found that, if the deepest dip (around day 800) is caused by comets, they must be relatively large (∼100 km). Furthermore, E. H. L. Bodman & A. Quillen (2016) concluded that, if the family of comets scenario is correct, it is likely that a planetary companion is needed to form these sun-grazing comets around KIC 8462852.

★ E-mail: cristina.madurga-favieres@warwick.ac.uk

The presence of a companion has become more plausible as various fits (F. Kiefer et al. 2017; F. J. Ballesteros et al. 2018; R. Bourne, B. L. Gary & A. Plakhov 2018) have shown that such a body could be consistent with the irregularities of the *Kepler* light curve. For example, a planet at 2.1 au from the star and an orbital period of ∼900 d that hosts a wide ring system surrounding it (one outer and inner ring) could explain some of the photometric dips (F. Kiefer et al. 2017). The presence of a cloud of Trojan asteroids orbiting its L5 orbital point provides a different and plausible origin of the light-curve dips (F. J. Ballesteros et al. 2018). For example, the planetary ring system would be associated with the big dip (D800), and the cloud of Trojans would cause the irregular series of dips (D1500). This would result in an orbital period of the planet-ring-Trojans system of ∼12 yr.

Along with the *Kepler* data, ground based observations have also been used to shed light on the behaviour of the star, and have reinforced the idea of an object transiting with a certain period. A 1572-d periodicity was found to be consistent with a subset of dips (one occurred in 2013, observed by *Kepler*, and another occurred in 2017, observed at Las Cumbres Observatory, LCO) (G. Sacco, L. D. Ngo & J. Modolo 2018). These post-*Kepler* dips, four of them of 1–5 per cent and lasting from several days to weeks, are consistent with extinction by optically thin dust clouds (T. S. Boyajian et al. 2018), as multiband photometry shows differential reddening. In contrast, there were no constraints on the colour of the long-term dimming, suggesting the short-term and long-term variations are produced by dust of different sizes and therefore may be caused by independent processes. It was previously suggested the long-term secular dimming could be due to some form of circumstellar material, made of larger dust grains than the dust responsible for the short-term variations (H. Y. A. Meng et al. 2017). *V*- and *g'*-band observations from 2017 May to November at the Hereford Arizona Observatory, HAO, showed a series of complex dimming events. These data were modelled simultaneously with *Kepler* photometry using a transit fit and both the D1500 *Kepler* dip and one main dip of the HAO series were interpreted as the result of the passage of a brown dwarf (BD) surrounded by a ring system in a 1600-d elliptical orbit (R. Bourne et al. 2018). The brown dwarf could have orbiting icy moons that would sublimate near periapsis, as comets would, generating the series of irregular dips.

Given all the exciting work previously done, we analyse the *Transiting Exoplanet Survey Satellite* (*TESS*) photometry data to potentially confirm or refute previous scenarios. We summarize the observations of Tabby's star considered in this work in Section 2. We show the analysis made to estimate the orbital period and mass of a potential companion in Section 3, using the archival *TESS* photometry data (including the new transit), as well as our own radial velocity (RV) observations. We also show the study done of the symmetry of the transit, and the modelling of it to a combination of exocomets. In Section 4, we show our new LCO data we use to rule out several orbital periods and the final parameters of the companion given by a joint fit analysis done with all the data. Our final discussion of the nature of the companion and other possible scenarios are elaborated in Section 5, and finally Section 6 shows the summary of the work and future observations useful to unveil more characteristics of the companion.

## 2 OBSERVATIONS

### 2.1 The discovery of a single transit event in the *TESS* data

*TESS* (G. R. Ricker et al. 2015) observed Tabby's star (KIC 8462852, TIC 185336364) in sectors 14, 15, 41, 54, 55, 81, and 82 (Camera/CCD pairs: 2–4, 2–3, 2–4, 3–2, 3–1, 3–2, and 3–1, respectively). Sectors 14, 15, 41, 54, and 55 were observed in short-cadence mode (either 120 or 20 s) with Target Pixel Files (TPF) available, whereas sectors 81 and 82 were taken in longer cadence (160 s) Full Frame Images (FFI). To produce precise and accurate photometry for our analysis of Tabby's star we utilized the TPFED/FFIED pipeline (T. G. Wilson et al. 2026). This tool retrieves calibrated TPFs and FFIs with the default quality bitmask using TESSCUT (C. E. Brasseur et al. 2019) and performs noise-optimized aperture photometry with radii selected via minimization of the Combined Differential Photometric Precision (CDPP; J. M. Jenkins et al. 2010; J. L. Christiansen et al. 2012) noise after background correction and removal of Not-a-Number fluxes and flux errors. Systematic noise is removed from the extracted photometry by utilisation of the *TESS* co-trending basis vectors (CBVs) and engineering quanternion measurements for the camera/CCD pairs that Tabby's star was observed in. This has been shown to remove flux modulation due to spacecraft jitter, momentum dumps, and background scattered-light (Lightkurve Collaboration 2018; A. Vanderburg et al. 2019; L. Delrez et al. 2021).

We analysed our produced light curves of all *TESS* sectors (a total of 160 d) and found a single unreported transit event during sector 15, on 2019 September 3 (Figs 1 and 9). Solar system asteroids passing through the *TESS* Field of View may effect background estimates that can mimic transits in target flux light curves. We assessed the *TESS* Simple Aperture Photometry background light curves of Tabby's star and find no evidence of variations due to asteroid crossing events. The transit exhibits a long duration of ∼21 h with a depth of ∼1.1 per cent. This long duration transit has a symmetric, round shape.

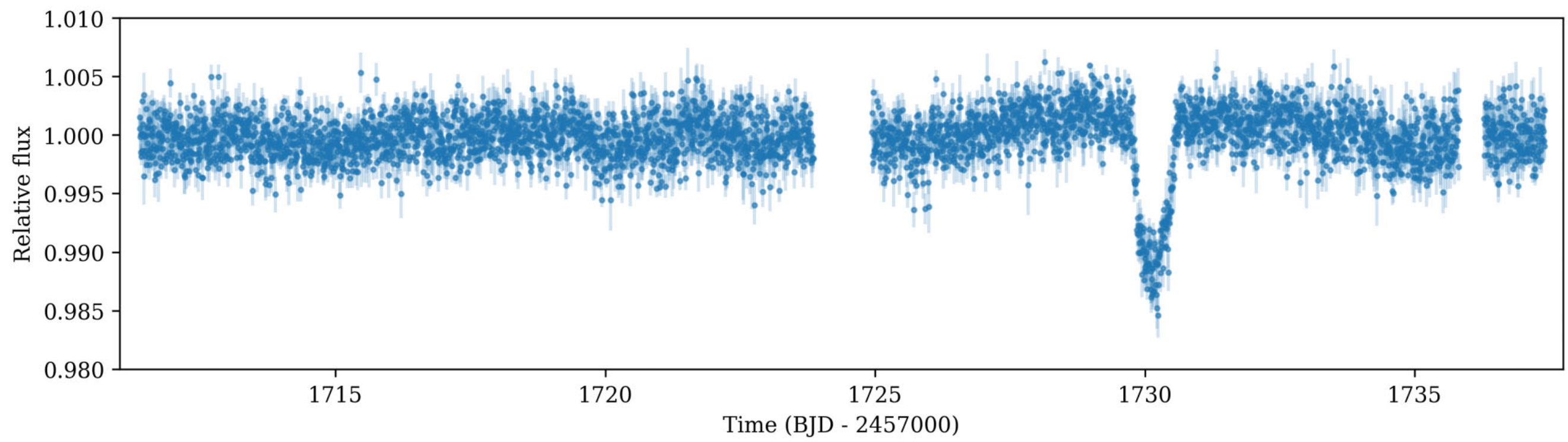

**Figure 1.** Sector 15 of the *TESS* data with the transit. It shows an increase of brightness before the transit occurs, which takes ∼5 d to reach its maximum, and another ∼5 d after the transit to disappear.

### 2.2 Radial velocity data

KIC 8462852 was previously observed spectroscopically with the intention of looking for spectral variations during the photometric dip events. The spectra were used to characterize the nature of the transiting dust (e.g. C. M. Lisse, M. L. Sitko & M. Marengo 2015; T. S. Boyajian et al. 2016, 2018; F. Kiefer et al. 2017; D. Lipman et al. 2019). Archival RV data of KIC 8462852 were taken with the FIES spectrograph at the Nordic Optical Telescope (NOT) (24 data points), the HARPS spectrograph on the ESO 3.6-m telescope (4 data points), the SOPHIE spectrograph at the Observatoire de Haute-Provence (OHP) (28 data points), and the HERMES spectrograph on the Leonard Euler telescope (16 data points). They cover the time period from 2015 to 2023 and thus bridge the gap between the *Kepler* and *TESS* data sets. The uncertainties in the RV measurements are generally large, with averages of $\sim 600\,\mathrm{ms}^{-1}$ for FIES, $\sim 300\,\mathrm{ms}^{-1}$ for HARPS, $\sim 730\,\mathrm{ms}^{-1}$ for SOPHIE, and $\sim 380\,\mathrm{ms}^{-1}$ for HERMES. The large uncertainties are caused by KIC 8462852 being a fast rotator ($v \sin i = 84 \pm 4\,\mathrm{km\,s}^{-1}$ T. S. Boyajian et al. (2016)), which leads to imprecise extraction of RV measurements from the standard data reduction pipelines due to the broadening of the cross-correlation function (CCF) caused by the rapid stellar rotation.

For the SOPHIE data, we performed additional analyses. First, we corrected the moonlight contamination following the method outlined in D. Pollacco et al. (2008), using the second SOPHIE fibre aperture, which observes the sky 2 arcmin away from the science target. This correction was systematically applied to all exposures. Next, instead of fitting the cross-correlation function with a simple Gaussian profile – as is typically done for slowly rotating stars – we modelled the CCF as the convolution of a Gaussian and a rotational profile (D. F. Gray 2021). This approach provided fits to the CCFs, but they were still noisy, broad and shallow. Due to the low quality of the spectra, the CCFs' uncertainties dominated over variations due to stellar activity. Activity indicators such as the full width at half maximum (FWHM) and bisector, which are highly sensible to the shape of the CCFs, were still not able to be robustly extracted. The FWHM of the Gaussian component, though not directly constrained, is expected to be small compared to the rotational broadening. We tested different approaches, fixing FWHM to a range of values, and also letting it vary freely. This method has already been successfully used to detect planets around fast-rotator stars, such as in A. Simonnin et al. (2025). We subsequently conducted the RV analyses and joint fit with the photometry data described in detail in Section 3.2 using these new time series. However, the planet parameter values did not change significantly (Table A1). We therefore fixed it to $7\,\mathrm{km\,s}^{-1}$ for all exposures, consistent with the SOPHIE instrumental profile in high-efficiency mode, which corresponds to about $7\,\mathrm{km\,s}^{-1}$. The projected rotational velocities were allowed to vary across exposures, yielding to a $v \sin i = 78.6 \pm 0.1\,\mathrm{km\,s}^{-1}$. This is consistent with the value $v \sin i = 84 \pm 4\,\mathrm{km\,s}^{-1}$. The measured RVs are derived from the fitted centre of the Gaussian and rotational profiles. This re-analysis of the SOPHIE data, resulted in a significant reduction of the average uncertainties to $\sim 130$ $\mathrm{ms}^{-1}$. This is why, in the end, only SOPHIE was used to ensure the most accurate results.

Motivated by this, further observations using OHP/SOPHIE were obtained as much as evenly distributed as possible in the period covering 2024 July to 2025 February through the OPTICON time allocation. 20 new data points were obtained, using high-efficiency and slow read-out mode. Each observation was taken with an exposure time of 1800 s and/or a maximum signal-to-noise ratio (S/N) of 30. Finally, we removed five out of the 48 total SOPHIE data points from our analysis because they were outliers or had high uncertainty. The 43 datapoints are shown in Table B1. During 2024 December 4 (3282 BJD-247000) several datapoints were collected attempting to measure the Rossiter–McLaughlin effect for a planet with a period of $P = 776$ d. For this period, there are two large absorption events in the Kepler data (D800 and close to D1500) which coincide with the *TESS* transit. The photometry data taken those days showed no transit (Section 3.3.1) and the RV variability on that day is comparable to the overall dataset (352 and $293\,\mathrm{m\,s}^{-1}$). Therefore, we attributed this variation to scatter. For our final analysis, we binned the data in order to achieve one data point per day. During 2024 December 6, six data points were taken, which we also binned. Therefore, the final RV data set contains 29 data points, shown in Fig. 2.

## 3 ANALYSIS

Given our serendipitous discovery of a seemingly symmetric and long-duration transit of KIC 8462852, we conducted a deeper analysis of the light curve and RVs to aim at establishing the cause of the observed transit.

### 3.1 Period estimate from a single transit

First, to understand the architecture of the system, we estimated the possible period by modelling the single transit event seen in the *TESS* data. This was conducted using the transit model parametrization and nested sampling technique implemented in the `Juliet` PYTHON tool (N. Espinoza, D. Kossakowski & R. Brahm 2019), with the use of the BATMAN package (L. Kreidberg 2015). We started our initial analysis with a uniform prior on the period from 776 to 2000 d. Periods shorter than 776 d can be ruled out by the *Kepler* data, as multiple transits would have been visible. The maximum value of 2000 d was conservatively chosen as a longer period is unlikely given the transit duration and the known stellar density, (T. S. Boyajian et al. 2016). The prior on the radius ratio was uniform from 0 to 0.13 (twice the radius of Jupiter given the stellar radius of KIC 8462852). This radius upper limit ensures we covered a radius that would correspond to the brown dwarf mass range up to the onset of hydrogen burning. As our photometry does not allow us to constrain the eccentricity, in this initial analysis, we fixed it at 0. We included the stellar density and a uniform distribution from 0 to 1 for the impact parameter. Our model includes a Gaussian Process (GP) regression with a Matérn kernel as it offers sufficient flexibility to model correlated noise effectively using only two hyper-parameters. The initial results report a period of $P_{\rm single} = 1103^{+242}_{-234}$ d.

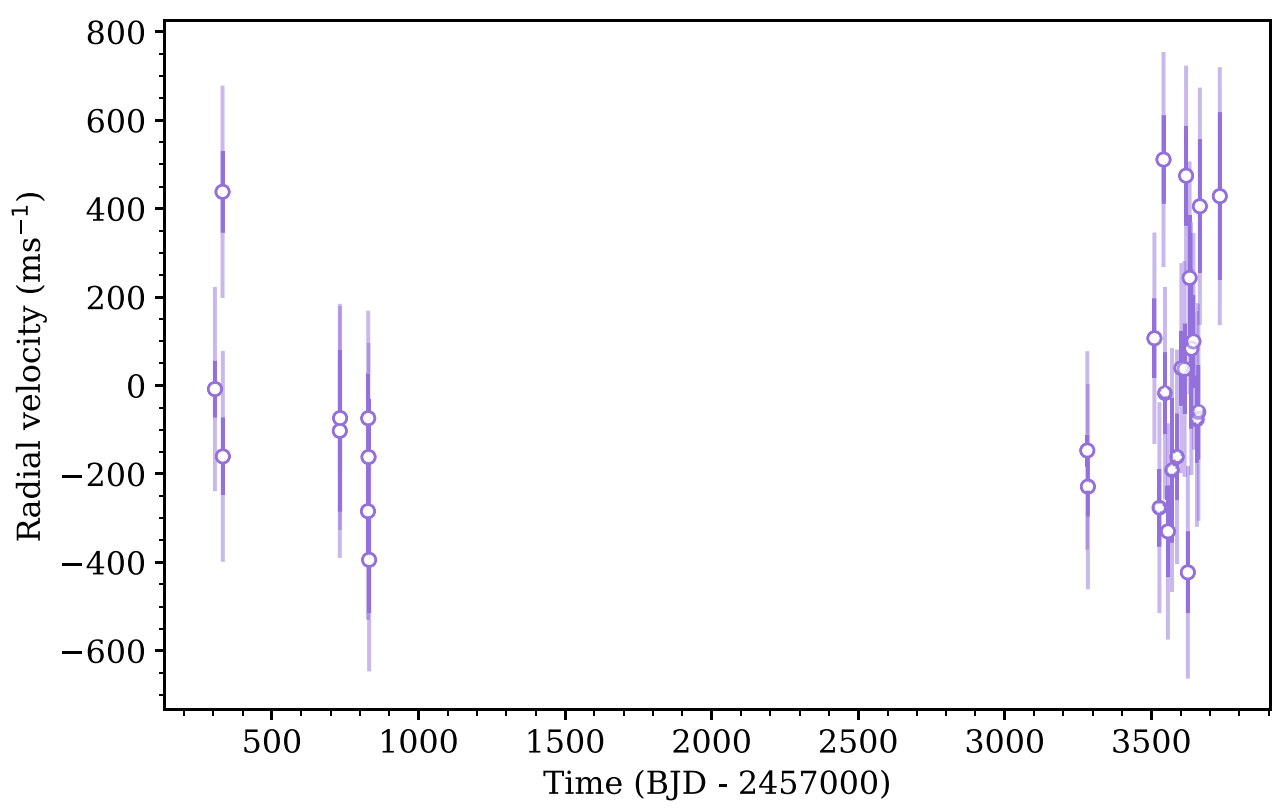

**Figure 2.** RV time series data taken at OHP, starting at 2015 October 11, and ending at 2025 February 27. The nominal, and the stellar jitter uncertainties are represented by the dark and light vertical lines, respectively.

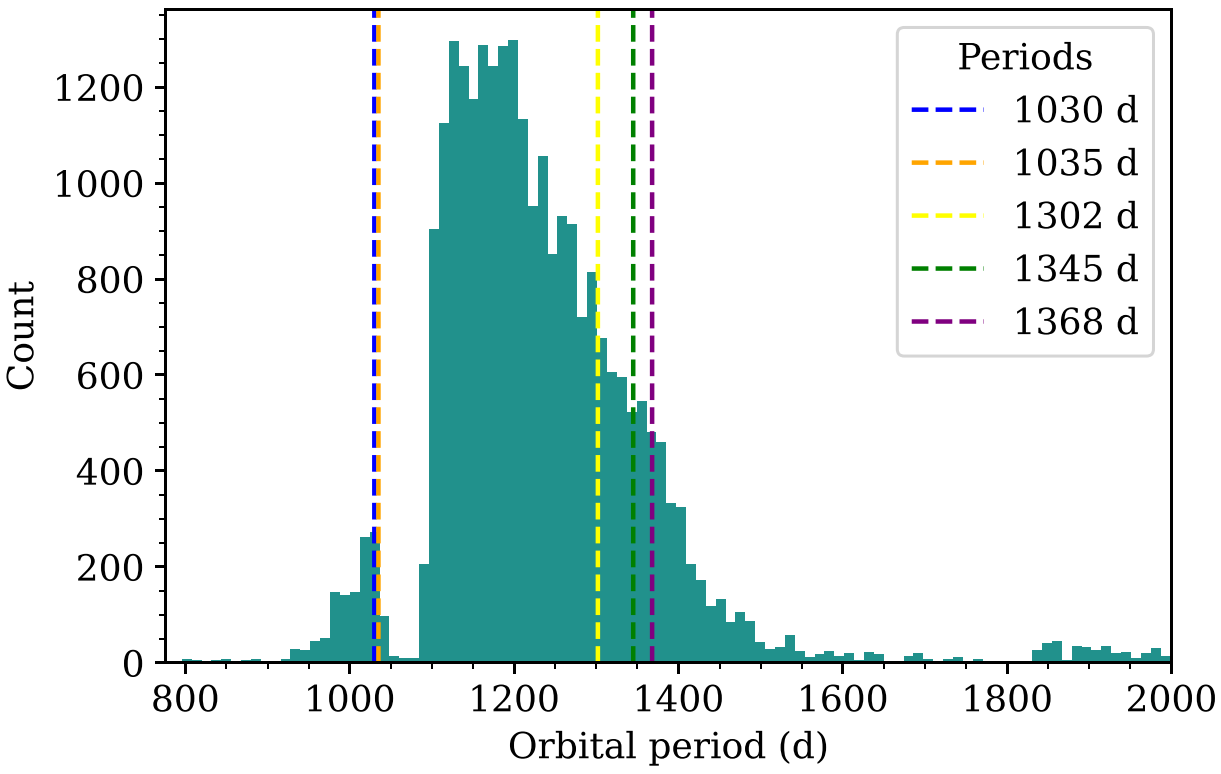

**Figure 3.** Posterior period distribution from *TESS* and SOPHIE data, showing the five possible windows constrained by *Kepler* (in detail in Section 3.3).

### 3.2 A joint fit using the *TESS* data and newly obtained radial velocity measurements

We obtained RV data with OHP/SOPHIE, and combined it with archival OHP/SOPHIE measurements. These were modelled jointly with the single *TESS* transit to refine the parameters of the transiting companion, in particular, to constrain the orbital period and estimating its mass from the fitted semi-amplitude. The priors used are later described in detailed in Section 4.2. We used the data from all *TESS* sectors (14, 15, 41, 54, 55, 81, and 82), but bin it to 30 min to greatly reduce the computational time. Given the long-duration and depth of the transit, this binning does not significantly affect the derived parameters. We used the JULIET package which has support for the MultiNest algorithm (F. Feroz, M. P. Hobson & M. Bridges 2009; J. Buchner et al. 2014) via the DYNESTY package (J. S. Speagle 2020) to explore the parameter space and generate posterior distributions. We performed our analysis with 2000 live points for the nested sampling calculations which ensures the entire parameter space (including our period range) was covered extensively. The photometric observations use the same Matérn kernel which we used in the initial photometry single transit fit.

To model the RV observations, we employed a Keplerian model that defines the semi-amplitude of the curve (REDVEL package; B. J. Fulton et al. 2018). We also performed several fits using an exp-sine-squared GP kernel alongside the Keplerian model, due to the potentially active nature of the star. Rapidly rotating stars such as KIC 8462852, (stellar rotation period of $0.8797 \pm 0.0001$ d from T. S. Boyajian et al. 2016) can show RV variations that can be driven by stellar activity (e.g. spots or plages). To account for additional noise coming from stellar oscillations, we implemented a jitter term (Section 4.2).

In our initial assessment of the OHP/SOPHIE and *TESS* data, we used different models with varying priors, and compared their Bayesian log evidences to ascertain the robustness of the results. In each fit, we further explored the analysis parameter space by changing the bounds of: the systemic velocity and the semi-amplitude of the RV data, and the GP hyper-parameters of the GP model only applied to the RV data. We ensured the the upper prior bound of the GP amplitude would cover the RV data scatter completely. In a further model, we removed the GP to test the impact of it on the RV data. We also fixed the semi-amplitude to 0 ms$^{-1}$ for both the analysis with and without the GP model. We applied all the different models to both all and the binned data. We finally chose the model with the highest log evidence, which was the one with no RV GP and a range of priors for the semi-amplitude values from 0 to 500 ms$^{-1}$, using the binned data. We found that the GP contribution is strongly non-dominant over the uncorrelated noise (62 ms$^{-1}$ over 215 ms$^{-1}$ the corresponding amplitudes), and therefore its use is not validated. This agrees with the fact that the model with no GP shows the highest log evidence. The joint fit returns a period $P = 1211^{+131}_{-85}$ d (Fig. 3). However, not all periods within this range are plausible, given the constraints imposed by other photometric surveys.

### 3.3 Constraints on the period from *Kepler* and other photometric surveys

With the orbital period posterior distribution from our joint fit, we used *Kepler* photometry to further constrain the possible orbital periods of the potential companion.

No other transit with the same depth or larger compared to the *TESS* transit, and the same duration, was found in the *Kepler*

**Table 1.** Transit windows and associated times.

| Window | Gap duration (d) | Transit start (UTC) | Transit end (UTC) |
|---|---|---|---|
| 1 | 3.5 | 2025-04-24<br>01:04<br>2460789.5 BJD<br>$P = 1029.7$ d | 2025-04-27<br>14:01<br>2460793.1 BJD<br>$P = 1031.5$ d |
| 2 | 1.6 | 2025-05-03<br>15:57<br>2460799.2 BJD<br>$P = 1034.5$ d | 2025-05-07<br>06:21<br>2460800.8 BJD<br>$P = 1035.3$ d |
| 3 | 5.1 | 2026-10-18<br>16:25<br>2461332.2 BJD<br>$P = 1301.0$ d | 2026-10-23<br>18:21<br>2461337.3 BJD<br>$P = 1303.6$ d |
| 4 | 0.0 (23 min) | 2027-01-15<br>07:47:02.458<br>2461419.8 BJD<br>$P = 1345.648 \pm 0.008$ d | 2027-01-16<br>08:15:50.458<br>2461421.8 BJD |
| 5 | 2.0 | 2027-02-28<br>06:21<br>2461464.8 BJD<br>$P = 1367.3$ d | 2027-03-02<br>06:20<br>2461466.8 BJD<br>$P = 1368.3$ d |

data, so we made the assumption that a previous transit event of this companion may either have occurred during a gap in the *Kepler* data or during one of the deeper photometric dips, which could conceal a transit. This assumption significantly constrains the possible period ranges due to the long, continuous observation baseline of *Kepler*. Any photometric event deeper than 0.1 per cent is masked in the *Kepler* data leaving instead a gap where the flux dip was. We chose to do this as these events would otherwise bias an analysis towards period and transit depth solutions centred on these typically deep photometric events instead fitting these events with a GP in the process. This is driven by the greater photometric precision of *Kepler* compared to *TESS*. A deeper analysis of the *Kepler* dips simultaneously with the *TESS* photometry containing numerous transit models would be very computationally expensive and is beyond the scope of this work.

We find that only five transit windows remain viable (Table 1), which fall in the period distribution as shown in Fig. 3. The most probable period corresponds to window 3 (yellow dashed-line). All other windows appear less likely but windows 1 and 2 will take place earlier, which facilitates future observations. A representation of when the possible transits have happened and will happen in the future is shown in Fig. 4. The phase-folded data for the five potential periods from *TESS* (sector 15) and *Kepler* data are shown in the Fig. E1. This illustrates the phase distribution of the dips in the *Kepler* data.

#### 3.3.1 *Looking for past transits in archival data*

We searched through archival data from other photometric surveys to see if a transit has been observed previously. We looked at photometric data from the ZTF (Zwicky Transient Facility), SWASP (Super Wide Angle Search for Planets), the LCO 0.4-m telescope network consisting of TFN (in Tenerife, Spain), and OGG (Hawaii, USA), TiMo (Titan Monitor at Lowell Observatory, USA), the AAVSO (American Association of Variable Star Observers), and the Hereford Arizona Observatory (HAO). These light curves have average flux errors ranging from 0.0015 to 0.0122 which could allow for detection of the *TESS* observed transit. None of the retrieved data show any clear transit nor any sufficiently strong variation to confirm a decrease in flux caused by a companion of the characteristics found in the *TESS* data and described above. Therefore, no archival data helped rule out the possible periods found with the *TESS*, the *Kepler* and the SOPHIE data.

### 3.4 Constraints from direct imaging

From direct imaging, it is known that KIC 8462852 has a faint (M2V) companion with a separation of ∼1.95 arcsec, which translates to a projected distance of ∼885 au (T. S. Boyajian et al. 2016), and a mass of $M = 0.44 \pm 0.02$ M$_\odot$ (L. A. Pearce et al. 2021). This companion is not sufficiently close to KIC 8462852 to be able to produce a transit. The magnitude difference between KIC 8462852 and the nearby M dwarf is $\Delta m = 5.4$ mag $\geq 2.5 \log_{10}(t_{12}^2/(t_{13}^2\,(R_c/R_*)^2))$ (S. Seager & G. Mallén-Ornelas 2003; A. Vanderburg et al. 2019), which implies the M dwarf is not bright enough to have produced the transit. Apart from this wide binary companion, no other body has been detected. From the orbital period from the joint fit above, the potential giant planet or brown dwarf around Tabby's star would be located at a distance of $\sim$ 0.005 arcsec. At such a small distance, the companion could not be resolved by direct imaging, and therefore, its presence could not be confirmed or ruled out previously using this technique (T. S. Boyajian et al. 2016; L. A. Pearce et al. 2021).

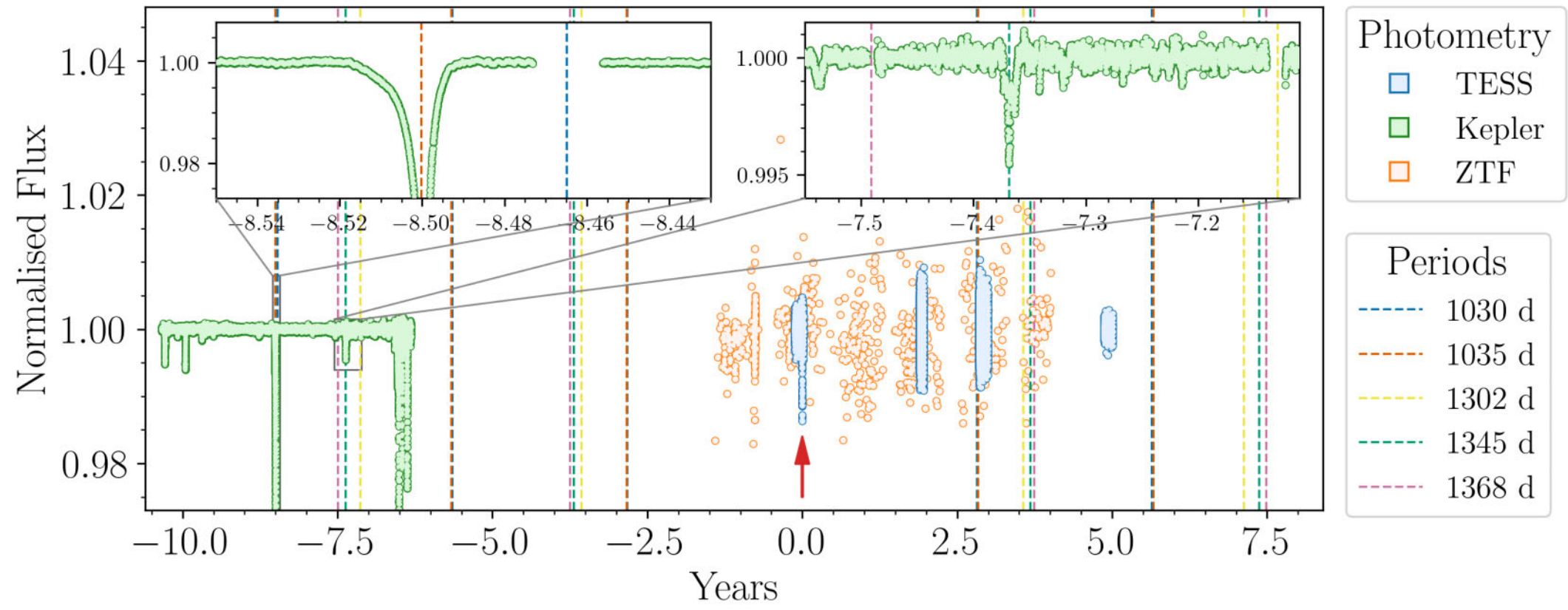

**Figure 4.** *Kepler*, *TESS*, and ZTF photometetric observation relative to the single *TESS* transit indicated with an arrow. The dashed vertical lines show the possible transit windows for various periods. The possible period solutions of ∼ 1030 and ∼ 1035 d fall right before sectors 54 and 55, which took place around 2.8 years after the transit. We show the ZTF data, as they were taken at the same time as some of the possible windows.

### 3.5 Exploring an alternative origin of the transit. Modelling a chain of exocomets

Despite of this previous analysis, the presence of the companion was still uncertain. With just one single transit, and the high scatter of the RVs, we were not able to currently conclusively confirm that both variations were caused by the same object. As the explanation of the irregularities of the system is, so far, a family of exocomets, we decided to study the shape and symmetry of the transit, to rule out it was caused by them. To address the possibility that it was caused by a combination of events, we conducted a series of analysis by modelling it as a chain of exocomets. We used two different models to fit the exocomets, and a Markov-Chain Monte Carlo (MCMC) algorithm to fit the transit data with them. The first fits were done using the exocomet model from A. Lecavelier des Etangs et al. (2022) (which we refer here as Model 1). The model states that the decrease in relative flux is:

$$\Delta F/F(t) = K \times (e^{-\Delta} - e^{-\Delta'}) \quad (1)$$

where $\Delta = \beta(t - t_0)$ if $t \geq t_0$, $\Delta = 0$ if $t \leq t_0$ and $\Delta' = \beta(t - t_0 - \Delta t)$ if $t \geq t_0$, $\Delta' = 0$ if $t \leq t_0 + \Delta t$. The four free parameters are shown in Table D1. For each fit, we defined the number of exocomets and allowed each of their shapes to differ. However, to achieve a good model fit and convergence of parameters, it was necessary to limit their priors to a small range, resulting in the exocomets becoming similar to one another. The priors of $t_{\rm beg}$ were set so that the exocomets were equally spaced during transit to mimic an exocomet chain and aid in fitting convergence, with each one exocomet starting to transit slightly after the previous one. $\Delta$t represents the time taken for the coma to transit. Therefore, one exocomet would cover the entire transit if $\Delta t = t_{\rm tr}/2$ (when ingress and egress are of equal length), where $t_{\rm tr}$ is the transit duration. We verified that, for more than one exocomet to transit, the value of $\beta$ had to be less than one for the MCMC method to converge. Finally, as the observed absorption depth of the transit is $0.012 \times 10^{-4}$, we set this as the upper limit of $K$ (with an uncertainty range). For each MCMC, we fitted the transit to one to ten, and then 20 exocomets. Fitting more would have been computationally expensive and 10 and 20 exocomets already produced similar fits. These number were consistent with F. Kiefer et al. (2017), who modelled Tabby's Kepler transit signatures using five and seven exocomets. We used the Bayesian Information Criterion (BIC) values to defined the best-fitting model, which used three exocomets (Fig. D1). Although a model with ten exocomets also provided a good fit, this model and all models requiring more than one exocomet are physically unrealistic, as they require a very low $\beta$ value (Table D1). We suspect that this is because such very low $\beta$ values produce essentially flat exocomet transit models that the fitting prefers to model the observed symmetric transit (Fig. D2). The lowest $\beta$ value found for exocomets is 0.97 d$^{-1}$, around $\beta$ Pictoris (A. Lecavelier des Etangs et al. 2022) more than two times bigger than the highest $\beta$ we obtained with our models. The model with one exocomet was the only one with a realistic $\beta$ value, but it resulted in a poor fit, especially during the egress. Furthermore, just one exocomet would not be capable of producing a dip of this depth.

**Table 2.** Priors and posteriors of the parameters of the best-fitting exocomet model for *TESS* and *Kepler* transits using Model 2. For the *TESS* fit, the x value represents the mid-transit time and increases by 0.3 for each exocomet, starting at $t_0 - (t_{\rm tr}/2)$; for the *Kepler* fit, it increases by 2. The transit duration is $t_{\rm tr} = 0.0058$ d for the *TESS* transit, and $t_{\rm tr} = 0.82$ d for the *Kepler* one.

| | $t_0$ (BJD-2450000) Mid-transit time | A ($10^{-4}$) Absorption depth | $\theta$ (d) Ingress | $\lambda$ (d) Egress |
|---|---|---|---|---|
| ***TESS*** | | | | |
| Prior | (x, x+0.3) | (10, 100) | (0, 1) | (0, 0.5) |
| Posterior | 8729.91 ± 0.01 | 70 ± 10 | 0.07 ± 0.01 | 0.20 ± 0.03 |
| | 8730.22 ± 0.01 | 80 ± 10 | 0.2 ± 0.02 | 0.10 ± 0.02 |
| | 8730.48 ± 0.01 | 80 ± 10 | 0.3 ± 0.03 | 0.04 ± 0.01 |
| ***Kepler*** | | | | |
| Prior | (x, x+2) | (10, 100) | (0.2, 3.5) | (1, 5) |
| Posterior | 6340.8 ± 0.2 | 12 ± 1 | 0.39 ± 0.07 | 2.8 ± 0.3 |
| | 6342.2 ± 0.3 | 13 ± 4 | 0.58 ± 0.23 | 2.8 ± 0.2 |
| | 6344.6 ± 0.1 | 80 ± 10 | 1.40 ± 0.08 | 2.3 ± 0.1 |
| | 6345.3 ± 0.1 | 40 ± 10 | 1.50 ± 0.10 | 1.3 ± 0.6 |
| | 6348.94 ± 0.03 | 12 ± 1 | 0.23 ± 0.02 | 2.7 ± 0.3 |
| | 6349.2 ± 0.4 | 12 ± 1 | 1.70 ± 0.05 | 1.1 ± 0.3 |

A second set of fits were performed with the asymmetrical comet transit model from G. M. Kennedy et al. (2019), Model 2, which defines the transit as a three-parameter Gaussian model before the transit centre; and with an exponential tail beyond it:

$$x_{\rm comet} = \begin{cases} 1 - A\exp\left[-\frac{(t-t_0)^2}{2\theta^2}\right] & t \leq t_0 \\ 1 - A\exp\left[-\frac{t_0-t}{\lambda}\right] & t > t_0 \end{cases} \quad (2)$$

The prior distributions of each parameter were set after an intensive survey in which we analysed the model with different parameters to narrow down the range of possibilities and optimize the MCMC (Table 2). We performed the MCMC with one to ten exocomets and obtained the best-fitting model with three exocomets (same as with Model 1) (Fig. 5). It should be noted that the fits were biased by the timing of each exocomet transit. Neither model produced very accurate or physically realistic fits that support the theory that the *TESS* transit was caused by a chain of exocomets.

On the other hand, we fitted a seemingly symmetric Kepler transit of Tabby's star to check if the transits could all be associated with exocomets. We fitted the transit at around 1511 BJD-2454833 using Model 2. After screening several values and testing them, we changed the priors to adjust to the characteristics of the transit (for example, its much longer duration compared to the *TESS* transit). We modelled it with one to eight exocomets, and obtained the best-fit with six exocomets (Fig. 5 and Table 2). As previously, the fits depended on the chosen transit time of each exocomets. We run models varying this, using exocomets that would take one, two and three days to transit. The best-fit was obtained with two days.

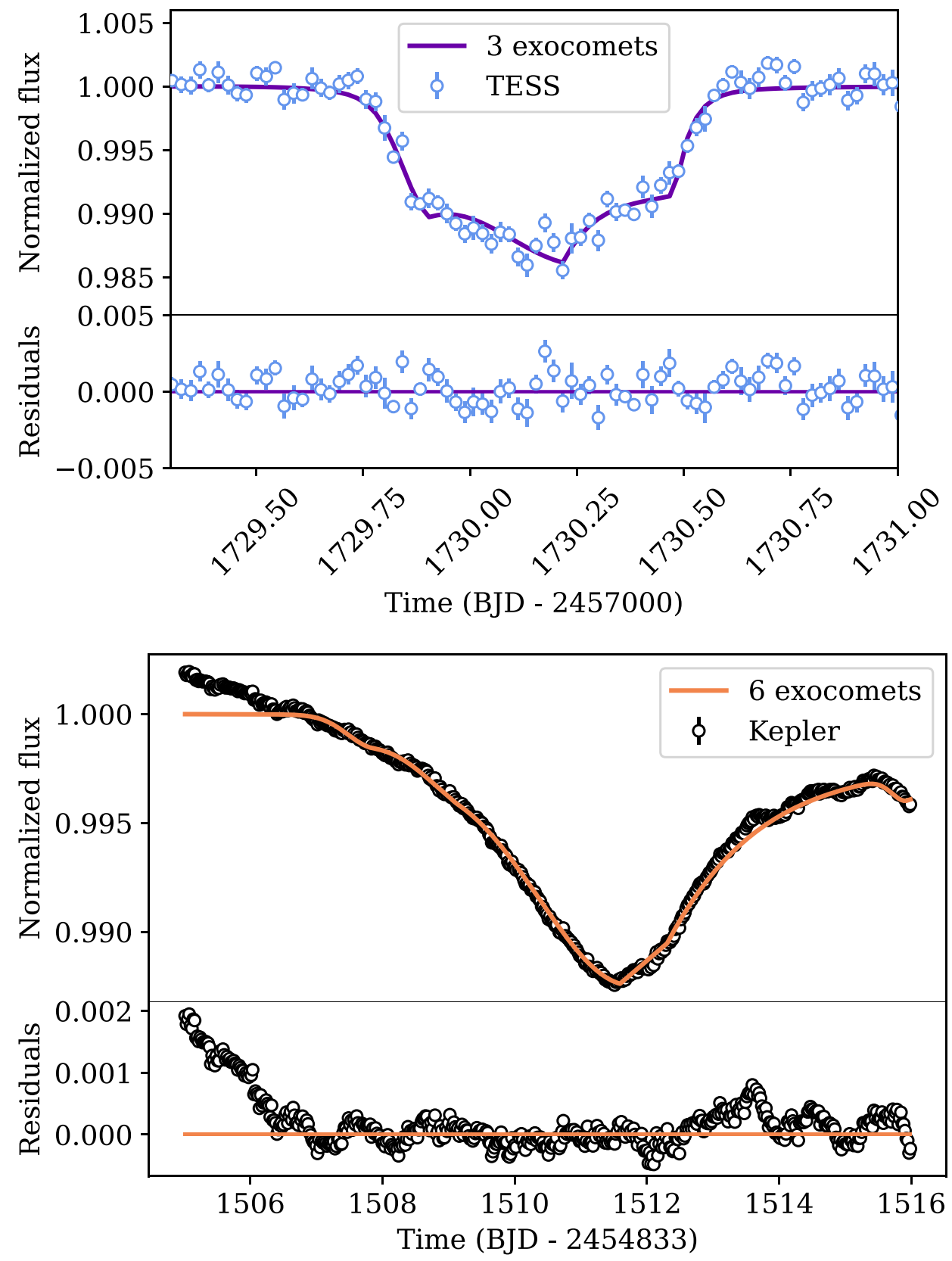

**Figure 5.** Top: *TESS* transit and best-fitting exocomet model with three exocomets using Model 2, with the residuals. Bottom: Kepler dip and the best-fitting exocomet model with six exocomets, again using Model 2, with the residuals. The fit agrees with the data.

### 3.6 Symmetry of the transit. Bisector analysis

We performed a bisector analysis to check the symmetry of the transit and to determine whether it is more probable that a rounded symmetric object caused the transit rather than a dust cloud or asymmetric objects. We performed a Gaussian Process Regression (GPR) to weight the uncertainties, using a squared exponential kernel and a grid of 100 data points. We obtained the bisector by using 20 different flux values (Fig. 6). It showed a deviation from a straight line of ∼0.00001 phase, which we attributed to scatter caused by noise. We believe the transit shows no pronounced asymmetric features that would suggest it was caused by several objects with asymmetric features, such as exocomets, or by an irregular object.

We repeated the process for the Kepler dip and found a more asymmetric bisector, mainly at the minimum and maxium flux values. In Fig. 6, we can appreciate the differences between the two bisectors. Between flux ∼0.990 and ∼0.998, the bisector from *Kepler* moves away from the middle to the right, characteristic of a transit with a different ingress and egress. Between flux ∼0.999 and ∼0.995 the deviation is much stronger.

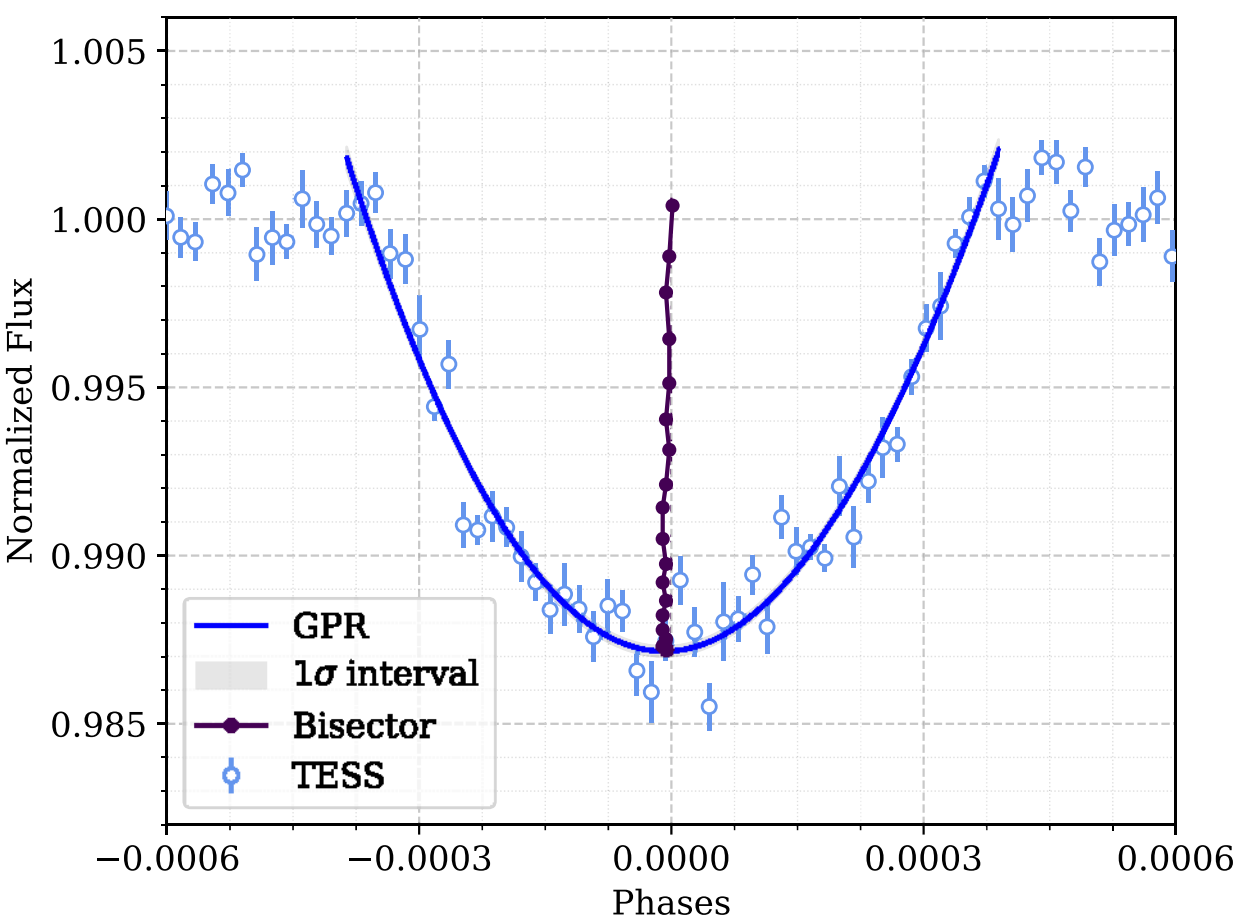

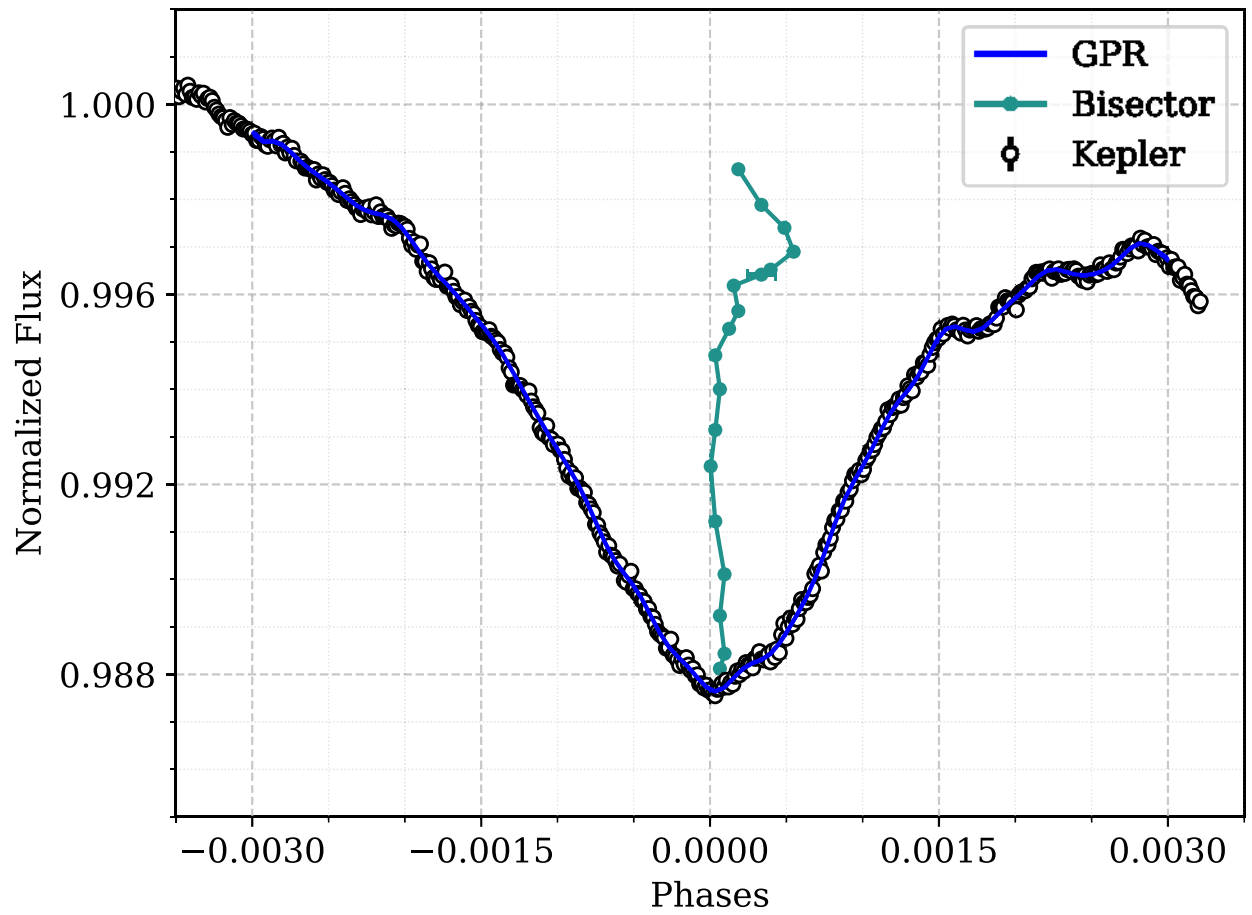

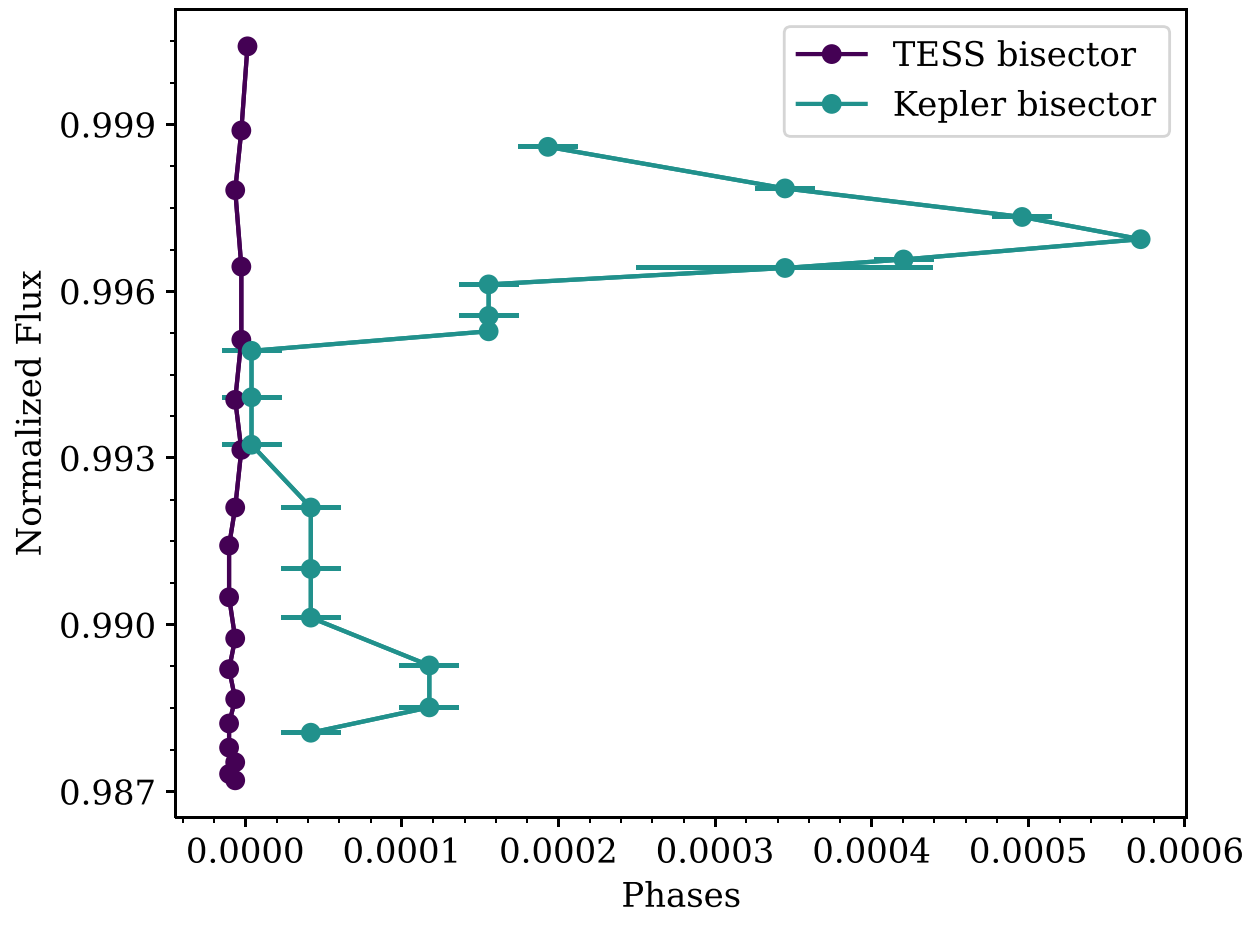

**Figure 6.** Top: *TESS* transit with its bisector. Middle: *Kepler* transit with its bisector. The uncertainties are smaller than the marker size. Bottom: Comparison between both bisectors.

## 4 RESULTS

### 4.1 New LCO photometry data during windows 1 and 2

To determine the orbital period of the potential companion, we secured photometric LCO time (PI: Madurga–Favieres) of Tabby's star during the first two time windows to attempt to observe a transit (Table 1). We used the 0.4-m telescopes in Teide, McDonald, and Haleakala observatories, allowing us to monitor the star as much as possible during the windows. Observations were divided into visits, each lasting around two hours. In the end, we had 13 visits during window 1 (25 h), and 10 visits during window

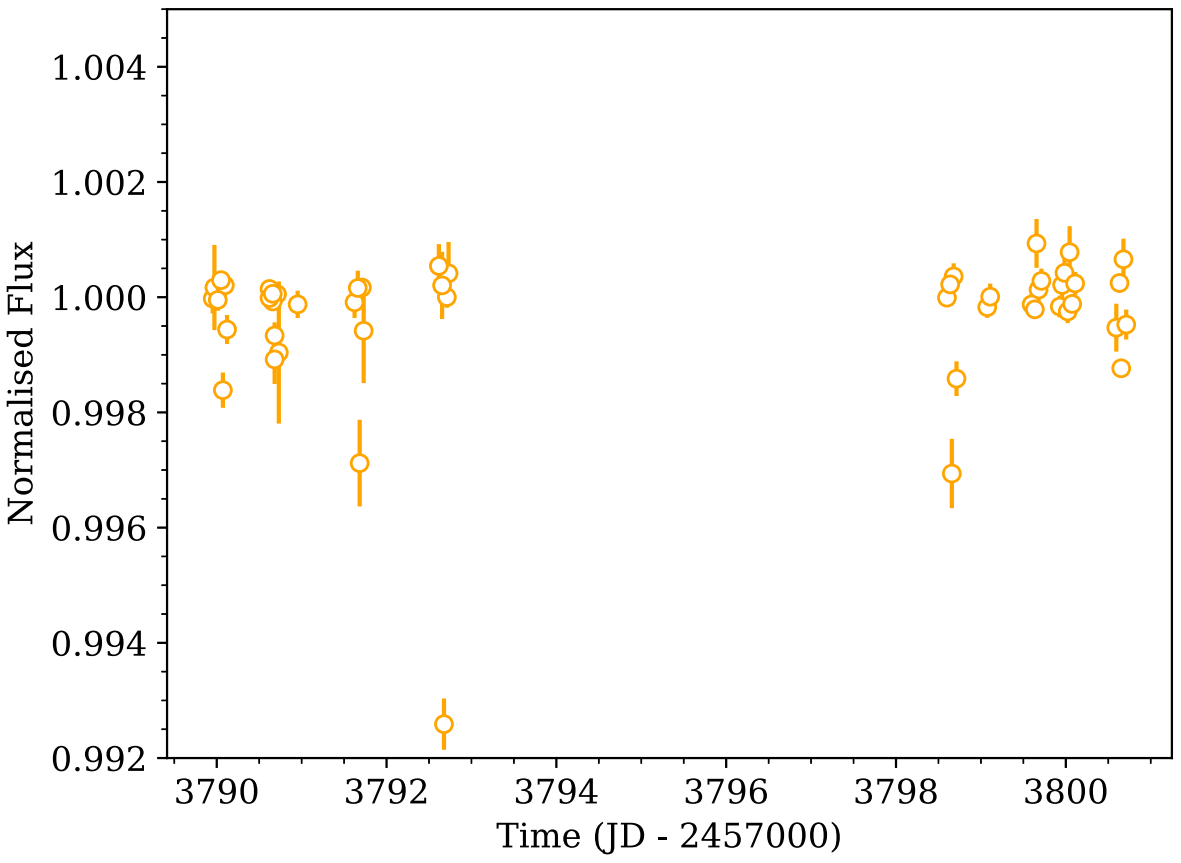

**Figure 7.** LCO data taken during windows 1 and 2, binned every 100 points (50 min).

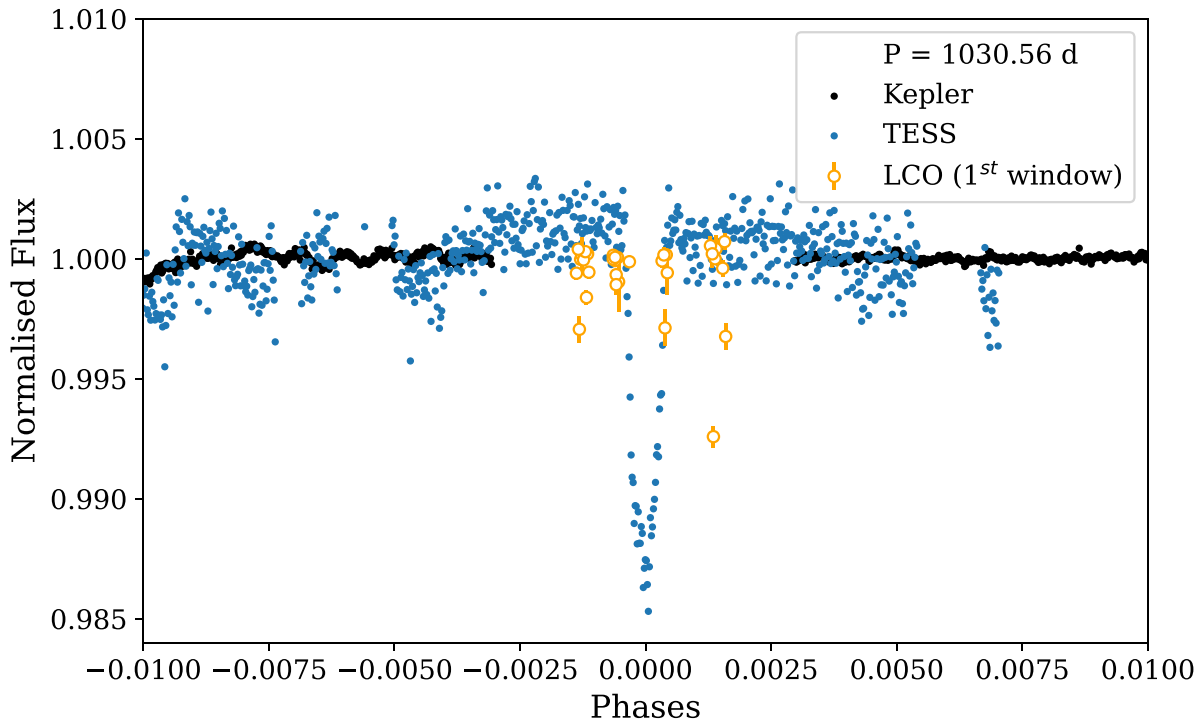

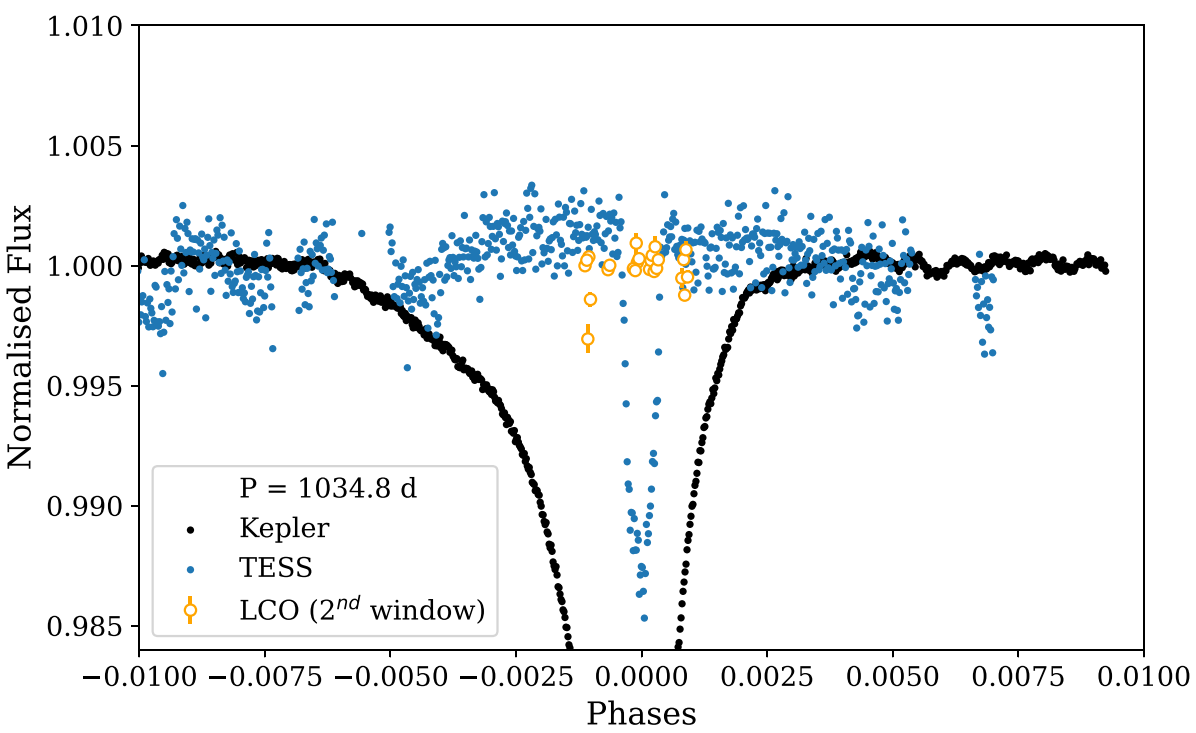

**Figure 8.** Top: Phase-folded *Kepler*, *TESS,* and LCO (window 1) data to a possible period. Bottom: Same as the top plot but with LCO (window 2) data phase-folded to a non-possible period.

2 (20 h and 50 min). Within each visit, each observation was taken with an exposure time of 30 s, using a gp filter. We used ASTROIMAGEJ software (K. A. Collins et al. 2017) to perform differential photometry to produce the light curves. The data of each visit were detrended individually using the airmass, the sky/pixel of the target star, and the comparison stars, with a combination of one, two, or three of them, choosing the subset that decreased the RMS the greatest. We finally binned the data every 100 points (50 min), which corresponded to one visit with one data point, seven visits with two data points, and eight visits with three datapoints (Table B2 and Fig. 7). With these data, we searched for transits to confirm or deny potential orbital periods of the companion. Due to weather conditions and network outages, not all the initial requested visits could be taken and we ended up with data gaps. Therefore, the data can be used to confirm or deny the existence of a new transit in them, but the gaps prevent us from learning the orbital period. For some periods, as shown in the left plot of Fig. 8, the gap matches the length of the transit, and therefore, a transit may have occurred during the gap. For some others, the LCO data rule out the possibility of a transit, as shown in the right plot of Fig. 8. Table 3 shows the final possible periods ranges from windows 1 and 2.

### 4.2 Results of the final joint analysis

Along with the new LCO data, we repeat the joint fit analysis. The obtained and derived parameters from the joint fit are shown in Table 4, as well as the priors. As it is a mono transit, the stellar density aids out period determination. It is derived using the stellar mass and radius values from the *TESS* Input Catalogue (K. G. Stassun et al. 2019): $M_* = 1.19^{+0.20}_{-0.15}$ M$_\odot$ and $R_* = 1.60^{+0.07}_{-0.08}$ R$_\odot$ The best-fitting solutions for the combined datasets are shown in Fig. 9, and the posterior parameter distributions in Fig. 10. The majority of the parameters follow well-constrained distributions, as expected. We can deduce that the companion would not be grazing because the impact parameter, $b = 0.76^{+0.05}_{-0.04}$, does not agree with (b $<$ 1 - $R_c/R_*$). This moderate value implies a U-shaped transit. The eccentricity is nominally low, and it is correlated with both the radius ratio and the impact parameter, preferring both a high impact parameter and radius ratio. We obtained a 2.3 $\sigma$ result for the mass of the companion, $M = 9.4^{+4.9}_{-4.1}$ M$_{\rm Jup}$ (3$\sigma$ upper limit $<$ 28.0 M$_{\rm Jup}$), and a radius of $R = 1.7 \pm 0.1$ R$_{\rm Jup}$. The mass agrees within 1$\sigma$ with the value obtained by the bootstrapping sine function fit used in Section C. As JULIET fit is more robust, we rely on this one more. We must point out that, as the mass is only a 2.3 $\sigma$ result, we cannot confirm or deny that the signal is due to the potential companion. We discuss this 2.3 $\sigma$ tentative mass detection in Section 5.

**Table 3.** Final possible periods ranges in windows 1 and 2.

| Window | Transit start $P =$ (d) | Transit end $P =$ (d) |
|---|---|---|
| 1 | 1029.65 | 1029.73 |
| | 1030.55 | 1030.64 |
| | 1030.91 | 1030.12 |
| | 1031.5 | 1031.52 |
| 2 | 1035.10 | 1035.12 |

## 5 DISCUSSION

### 5.1 The nature of the companion

As mentioned in Section 4.2, from our analysis, Tabby's companion would have a resultant mass of $M = 9.4^{+4.9}_{-4.1}$ M$_{\rm Jup}$ and a radius of $R = 1.7 \pm 0.1$ R$_{\rm Jup}$. Both parameters suggest the object would lie in the boundary region between a giant planet and a brown dwarf (left plot of Fig. 11). Nominally, this is 13 M$_{\rm Jup}$ that corresponds to the limiting mass for thermonuclear fusion of deuterium. However, this is an overlapping region between the two categories, and there is not a completely agreed limit between one and the other (J. Schneider 2018). For instance, based on formation scenario, the NASA Exoplanets Archive states the limit would be 30 M$_{\rm Jup}$. Giant planets appear to form by core accretion in a protoplanetary disc, while brown dwarfs would form similar to stars, by gravitational collapse of a molecular cloud (G. Chabrier et al. 2014). Moreover, J. Schneider et al. (2011) suggest that the transition between gas-giant planets and brown dwarfs occurs at much higher mass, around 25 $\mathrm{M_{Jup}}$. Independently of the fusion of deuterium, the IAU working definition of an exoplanet also states that exoplanets should have a mass ratio with the central object below the L4/L5 instability ($\mathrm{M/M_{central}} < 1/25$), no matter how they formed (A. Lecavelier des Etangs & J. J. Lissauer 2022). Under this condition, our companion enters the category of an exoplanet, as its mass ratio is $0.008 < 1/25$.

**Table 4.** Prior and posterior parameters values of the companion from the joint fit using the *TESS*, SOPHIE, and LCO data. The eccentricity priors are taken from V. Van Eylen et al. (2019) for systems with a single detected transiting planet. Regarding the period posterior, as shown in Section 3.3, only a subset of the identified five period windows that are within the quoted interval are compatible with the Kepler light curve.

| Parameter | Description | Prior | Posterior |
|---|---|---|---|
| *Transit* | | | |
| $P$ (d) | Orbital period | Uniform (776, 2000) | $1211^{+89}_{-78}$ |
| $t_0$ (BJD) | Mid-transit time | Normal (2458730.164, 0.005) | $2458730.164 \pm 0.004$ |
| $R_c/R_*$ | Radius ratio | Uniform (0, 1) | $0.111 \pm 0.004$ |
| $b = a\cos i/R_*$ | Impact parameter | Uniform (0.0, 2.0) | $0.76^{+0.05}_{-0.04}$ |
| $\mathrm{GP}_\sigma$ (ppm) | Amplitude of the GP | Log uniform (1e-6, 1e6) | $(32 \pm 2) * 10^{-5}$ |
| $\mathrm{GP}_\rho$ | Time/length-scale of the GP | Log uniform (1e-3, 1e3) | $0.80^{+0.17}_{-0.14}$ |
| *Radial velocity* | | | |
| $K$ ($\mathrm{ms}^{-1}$) | RV semi-amplitude | Uniform (0.0, 500) | $162^{+76}_{-69}$ |
| $e$ | Eccentricity | Beta (1.58, 4.4) | $0.09 \pm 0.05$ |
| $\omega$ (°) | Argument of periastron | Uniform (0, 360) | $141^{+84}_{-89}$ |
| $\mu$ ($\mathrm{ms}^{-1}$) | Systemic velocity | Uniform (0, 5000) | $4177^{+76}_{-63}$ |
| $\sigma_w$ ($\mathrm{ms}^{-1}$) | Jitter term | Log uniform (0.1, 1e5) | $212^{+40}_{-33}$ |
| $\rho_*$ ($\mathrm{kgm}^{-3}$) | Star density | Normal (409, 148) | $485^{+100}_{-96}$ |
| $M_c$ ($\mathrm{M_{Jup}}$) | Mass | | $9.4^{+4.9}_{-4.1}$ |
| $R_c$ ($\mathrm{R_{Jup}}$) | Radius | | $1.7 \pm 0.1$ |
| $\rho_c$ ($\mathrm{gcm}^{-3}$) | Density | | $2.2^{+1.3}_{-1.0}$ |
| $T_{eqc}$ (K) | Equilibrium temperature | | $268 \pm 13$ |

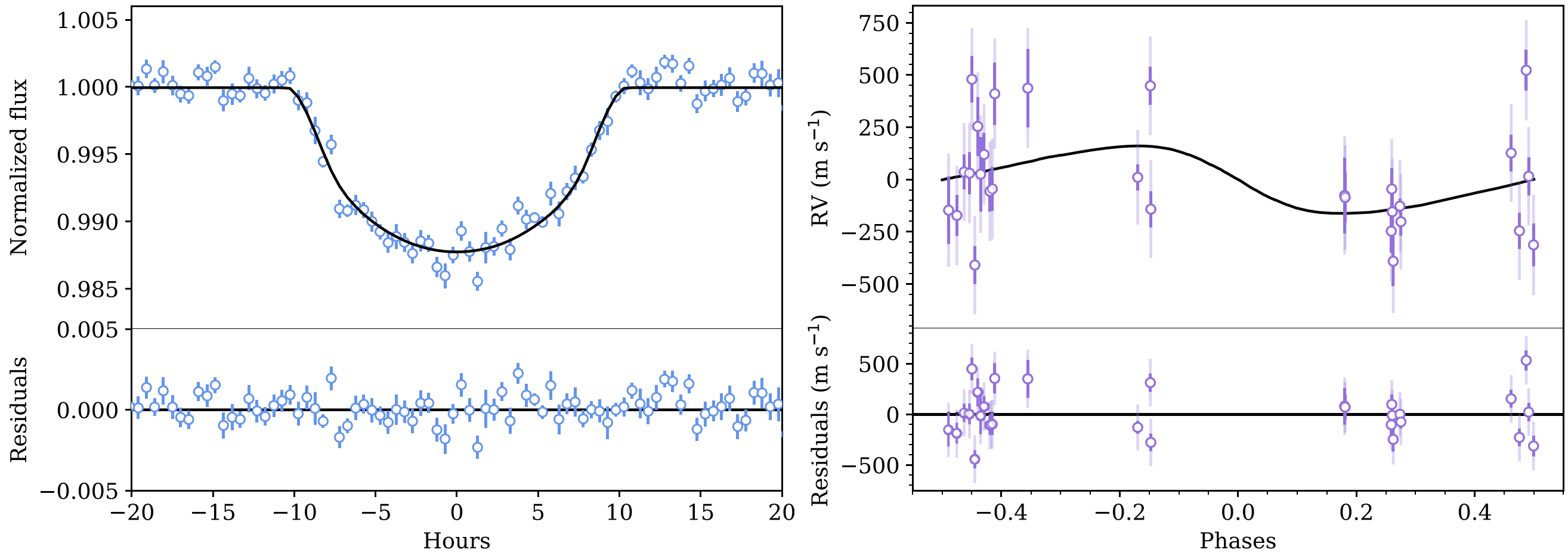

**Figure 9.** Left: Zoom in on sector 15 of the *TESS* data, showing the transit (dots) and the joint fit (solid line), and the residuals. Right: Phase-folded RV data (dots) fitted to the best solution from the joint fit (solid line), and the residuals.

The radius of the companion agrees with those of transiting brown dwarfs (typically between 0.1 and 1.6 $\mathrm{R_{Jup}}$; T. W. Carmichael et al. 2021; N. Grieves et al. 2021), and of giant planets. From self-compression (for giant planets) until the hydrogen burning phase (which defines stars), the mass–radius relation is given as $\mathrm{R} \propto \mathrm{M}^{-0.06}$ (S. Müller et al. 2024). Importantly, this does not differentiate giant planets from brown dwarfs.

From our mass and radius of Tabby's potential companion, we obtain a density of $\rho = 2.2 \pm 1.3\,\mathrm{g\,cm^{-3}}$, which agrees with those of the giants planets. Brown dwarfs on the other hand, tend to have larger densities, that can go from 10 to up to $1000\,\mathrm{g\,cm^{-3}}$ (for masses between 13 $\mathrm{M_{Jup}}$ and $\sim 85$ $\mathrm{M_{Jup}}$ in the right plot of Fig. 11). In particular, at the high end of their mass range, brown dwarfs become electron degeneracy supported objects (J. Pinochet 2019), leading to stop their contraction. The density of these type of objects is ~30 g cm$^{-3}$ (A. Odrzywolek & J. Rafelski 2018). This can lead us to think that the companion would be a very massive gas planet. Following this, Fig. 11 shows that giant planets and brown dwarf have similar behaviours in the mass–density relation, as suggested by S. Müller et al. (2024) or A. P. Hatzes & H. Rauer (2015). The latter pointed out that there are no differences in the mass–density relation between objects on each side of the 25 $M_{Jup}$ limit. In fact, to clearly differentiate brown dwarfs from the giant planets, they placed the limit between them at 60 $M_{Jup}$, when the electron degeneracy happens.

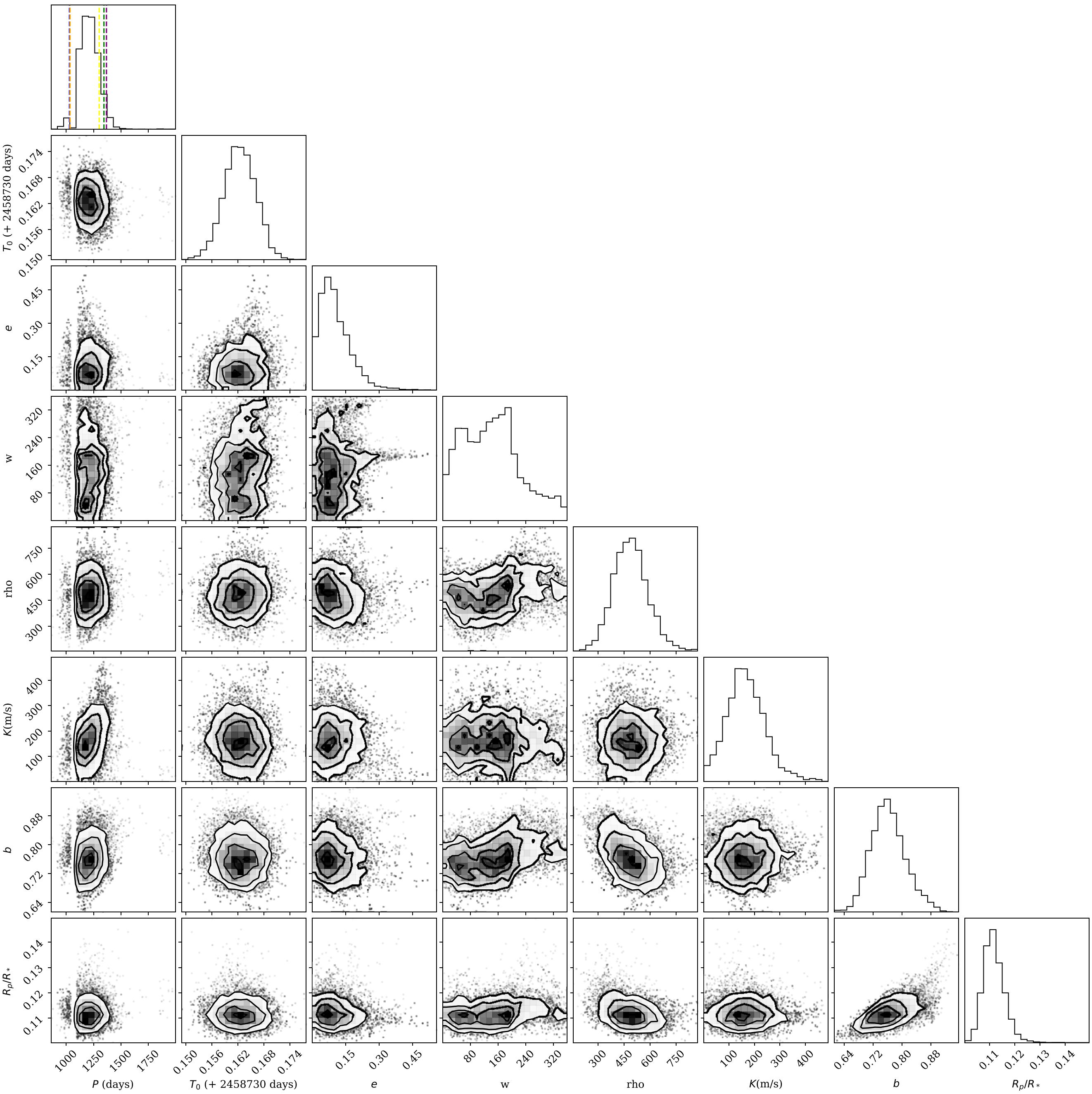

**Figure 10.** The posterior distributions of the parameters from the joint fit using the *TESS*, SOPHIE, and LCO data. The vertical dashed bars show in the period histogram indicate the midpoint of the remaining period solutions as imposed by *Kepler* data.

As the period is not yet defined, we have used $P = 1300$ d to compare it with the rest of the sample, as it is the most probable among the possible windows and subtle variations in period do not affect the conclusions given below. Tabby's companion would be one of the planets with the longest transit periods ever found, this is interesting as it is strongly related to the equilibrium temperature. The companion has an equilibrium temperature of $T_{eq} = 268 \pm 13$ K (assuming an albedo of $A = 0$), a low value compared to the rest of the sample due to its great distance from its host star (Fig. 12). The temperature of the companion agreed with those of the exoplanets with periods and masses similar to it (P. Sarkis et al. 2018; O. D. S. Demangeon et al. 2021; F. Philipot et al. 2023; G. Maciejewski et al. 2024; N. Grieves et al. 2025; N. Heidari et al. 2025). However, Tabby's companion, if confirmed,

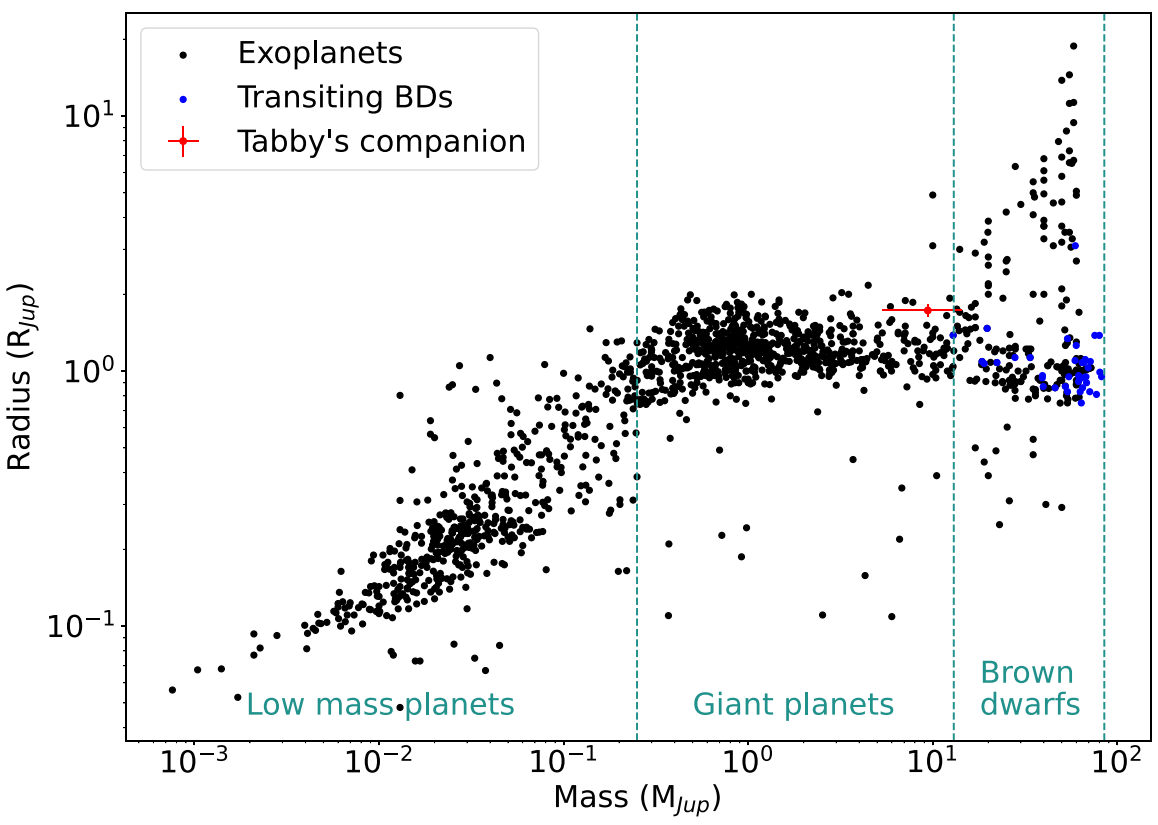

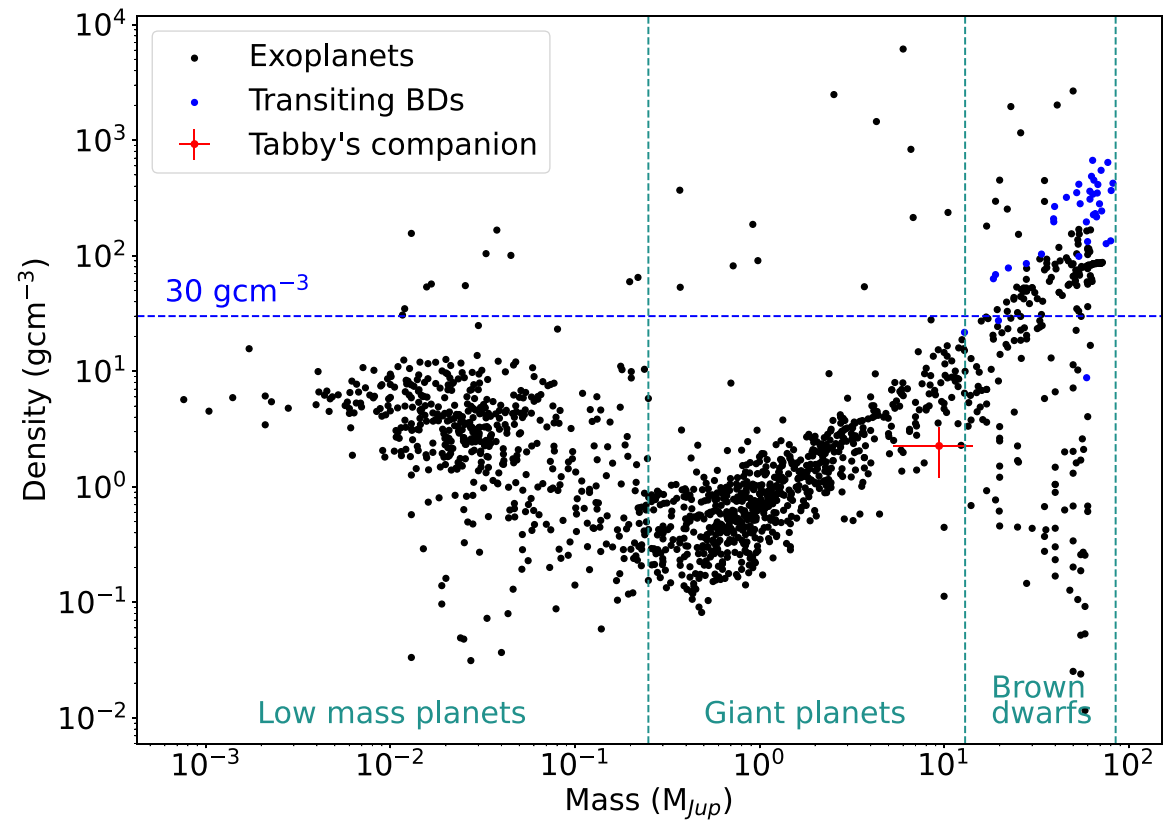

**Figure 11.** Planet radius as a function of planet mass (left) and planet density as a function of planet mass (right) for the exoplanets with uncertainties ≤30 per cent (from http://www.exoplanet.eu in 2025 February 7), and the transiting brown dwarfs (T. W. Carmichael 2023). The dashed lines are the limits between the mass regions (at 0.25, 13, and 85 $M_{Jup}$).

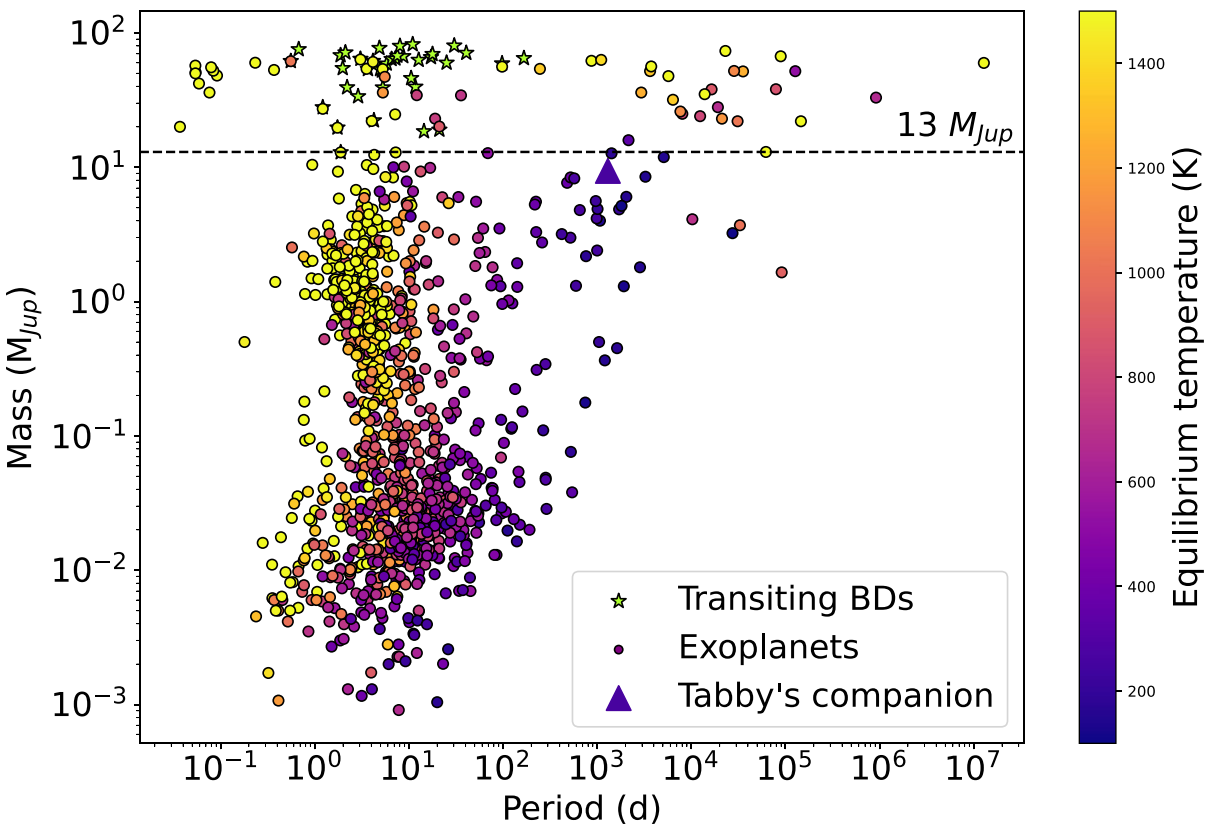

**Figure 12.** Planet mass as a function of the orbital period of the NASA Exoplanets Archive exoplanets (dots). They are coloured depending on their $T_{eq}$. For the transiting brown dwarfs (stars) we are showing the effective temperature $T_{eff}$. Their colours are outside the colour bar. Their temperatures are higher than that of the exoplanets, from 3100 to 9100 K.

would be the only one that has been seen transiting its host star, making it unique.

Its orbital eccentricity $e = 0.09^{+0.08}_{-0.05}$ hints at the period-eccentricity potential connection, where objects at large distances from their host star could have eccentric orbits. For brown dwarfs, their distribution of eccentricities differ significantly from that of long-period giant planets (at distances between 5 and 100 au from their host star; B. P. Bowler, S. C. Blunt & E. L. Nielsen 2020). This previous study found that giant planets tend to have low eccentricities, between $e \sim 0.05$ and 0.25, while brown dwarf companions show higher values, between $e \sim 0.6$ and 0.9. Although Tabby's companion would be at $a = 2.35 \pm 0.17$ au, it highly would agree with the giant planet distribution. For eccentric orbits, transits can be seen without occultation (J. N. Winn 2010), which would explain the lack of an occultation in the data.

Lastly, we attempted to determine the age of the system via gyrochronology. However, the stellar type of KIC 8462852 , F3V, is outside the range for known models (e.g S. A. Barnes 2007; R. A. García et al. 2014), which are for F7V or later type stars. In fact, the temperature range corresponding to spectral types F3V–F6V seems to represent the point at which the correlation between rotation and activity begins to break down (E. E. Mamajek & L. A. Hillenbrand 2008). If the companion were a brown dwarf, this would be more consistent with a young system, since the radius of brown dwarfs is strongly correlated to their age, as it decreases over their lifetime. If the system were young, this could cause the companion's radius to be inflated due to the residual heating, although this is speculative.

### 5.2 The possible existence of debris around the companion

The *TESS* data show a brightening centred on the transit (Fig. 1). A similar brightening is not seen after inspecting the other six sectors. If the brightening is astrophysical in nature, then it may be caused by forward scattering of dust particles around the companion (M. A. M. Kooten, M. Kenworthy & N. Doelman 2020). It may also be causing the transit event to appear deeper as both the companion and the surrounding dust would contribute to blocking the light from the star. The detection of a future transit could confirm or refute this hypothesis.

### 5.3 Other scenarios

Even though our data imply the presence of a large companion orbiting Tabby's star, other scenarios have been studied. We find that the observations do not support them, but we present them here as they cannot be completely ruled out.

#### *5.3.1 M-dwarf eclipsing binary*

We considered the transit to be caused by an M dwarf transiting the nearby M dwarf from Tabby's star, motivated by a ~0.4 per cent transit found in the *Kepler* data that agrees with the period of window 4 (Fig. 13).

Despite being less than half as deep as the *TESS* transit, it has the same duration. As *TESS* and *Kepler* observe at different wavelength rage, the different in depth could be due to differences in the stellar spectral energy distribution, and the transits could be caused by the same object. However, a transit produced by an eclipsing binary of two equal stars is V-shaped, unlike the more U-shaped ($b = 0.76^{+0.05}_{-0.04}$) transit in Tabby's star *TESS* light curve. This indicates that the transiting object is smaller than the host star. To obtain a U-shape, the companion would have to be as small as an M4 dwarf, or the impact parameter would have to vary to get an orbital inclination of $\sim 89.9^{\circ}$. In addition, eclipsing binaries produce two transits per orbit. We checked for the existence of a secondary transit with the same characteristics without success.

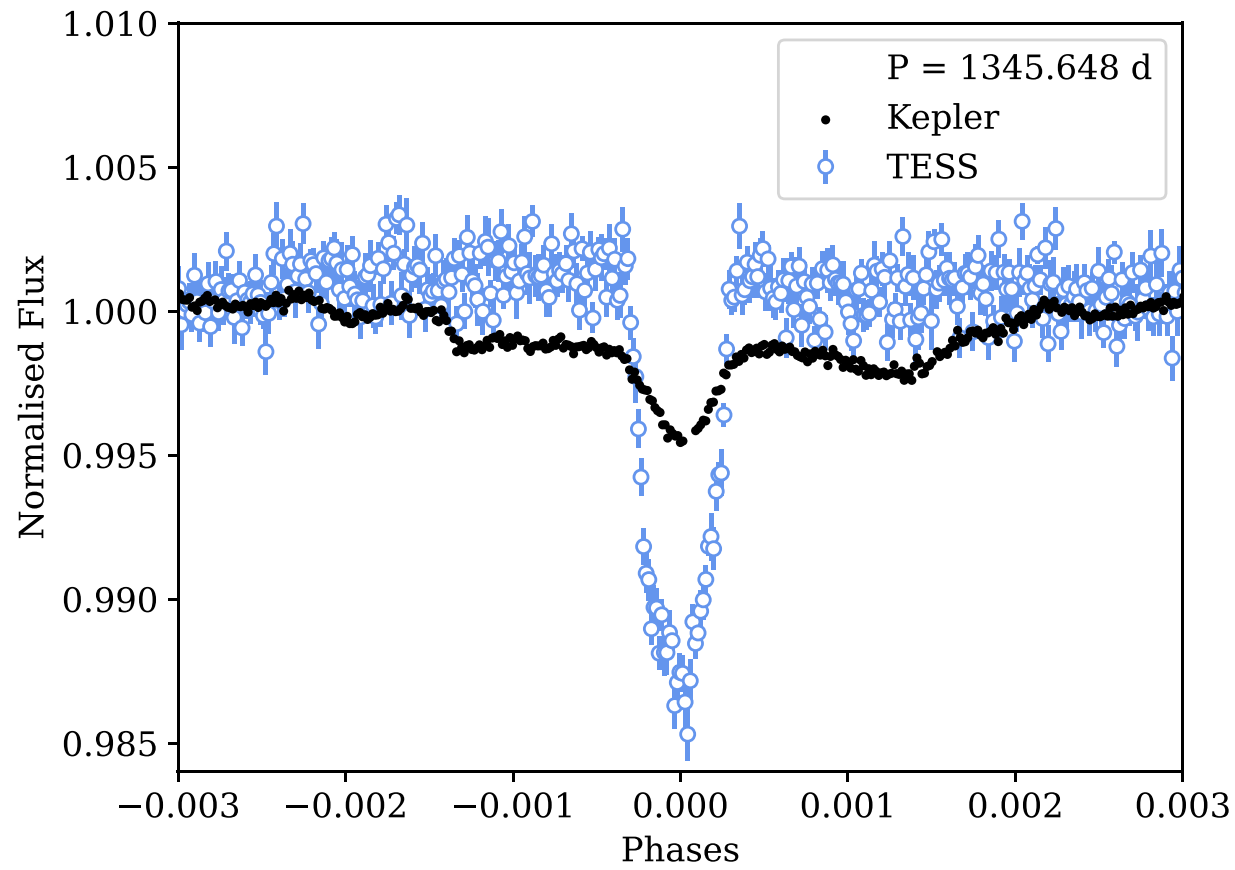

**Figure 13.** TESS transit (big dots with vertical lines) and Kepler transit (small dots) phase folded to the period of window 4.

We checked the *Gaia* Renormalized Unit Weight Error (RUWE) value associated with the nearby M dwarf. This value, which indicates how well a single-star model fits the observations, is expected to be around 1, and a significantly larger value could indicate that a star is a binary. For known single stars with G magnitude between 17 and 18 mag (KIC 8462852 B is 17.6 mag) with a five-parameter solution, the mean RUWE value is 1.010 (L. Lindegren et al. 2021a). In this case, the RUWE value of the M dwarf is 1.156, which is slightly higher. However, the astrometric excess noise of the M dwarf is 0.470, which completely agrees with the corresponding median astrometric excess source noise of single stars within the range of G magnitude (between 0.409 and 0.606). This does not support the eclipsing binary theory.

The *Gaia* photometry shows that the M dwarf is well situated in the M dwarf region of the HR diagram (*J*–*K* colour), which indicates that the potential companion to this M dwarf is not a star of identical type. If it were, the flux would be twice that of a single star, and it would deviate the star from its corresponding position in the HR diagram. However, we should be cautious, as the *J*–*K* colour is not sufficiently reliable due to systematic biases in *Gaia*'s parallax measurements (L. Lindegren et al. 2021b); and there is no *B*-*R* colour available.

All of these aspects argue against this scenario and the companion would more likely be a giant planet around Tabby's star rather than an eclipsing binary.

### *5.3.2 Dust cloud or debris*

Transits produced by debris or dust clouds are often irregular (B. P. Powell et al. 2021), which makes this scenario highly unlikely. Moreover, a dust cloud would not produce a non-zero RV signal as favoured by our fitting. As we analysed in Section 3.6, the transit presents a highly vertical bisector through the centre, with hardly any deviation from the vertical at phase = 0 (which would indicate that the transit is not symmetrical). Given its symmetry, it is highly unlikely that an object as irregular as a cloud of dust or debris could have caused it. As mentioned earlier, we believe that, if there is dust in the system, it would be around the companion, not on its own.

### *5.3.3 Chain of exocomets*

Despite exocomets being characterized by their asymmetric transit, they can sometimes transit in groups. When their transits are combined, they create a more symmetric transit (F. Kiefer et al. 2017; A. Etangs 2023). We studied this phenomenon (Section 3.5) using two different exocomets models, both of which struggled to fit the transit with a different number of exocomets. On the other hand, the fact that, we successfully fitted one of the most symmetric dips seen in the *Kepler* data, lends weight to the idea that this scenario is not favoured. We believe that this transit was caused by an object other than those causing the dips in the *Kepler* data.

## 6 SUMMARY AND CONCLUSIONS

This work presents a symmetric round-shaped transit found in the *TESS* data of KIC 8462852, suggesting the existence of a companion, either a giant planet or a brown dwarf, around the star. No transiting companion has ever been detected around this well-known star, so the potential evidence of a candidate presented in this work is significant, as its existence could explain the complexity of the system. The strongest theory is that a group of exocomets or planetesimal fragments are responsible for the irregular and apparently non-periodic dips in the *Kepler* light curve of Tabby's star. The presence of a companion would explain why these bodies are driven to the vicinities of the star, breaking up, as it would orbitally perturb them.

We have conducted an analysis to determine the period and principal parameters, such as the mass and radius, of the potential companion. The transit, found in sector 15 of the *TESS* data, has a duration of $\sim 21$ h and a depth of 1.1 per cent. To complement the photometry data, we have monitored the star during around six months to obtain new RV observations at OHP. A joint fit with all the *TESS* data (sectors 14, 15, 41, 54, 55, 81, and 82) and the new and archival RV data was done to estimate the period of the potential companion, which ended up being of at least $\sim$1030 d. We looked for a previous transit with a comparable depth and/or duration in the *Kepler*, ZTF, SWASP, LCO, TiMo, AAVSO, and HAO data, but none were found. We also made the assumption that the previous transit occurred during a data gap, or during one of the deeper dips of the *Kepler* data. This highly constrained the posterior period distribution from the joint fit to only five possible windows, around the periods: 1030, 1035, 1302, 1345, and 1368 d. We conducted LCO observations to follow-up the first two of them, searching for a transit to confirm or deny the potential periods of the companion. Due to some data gaps, not all periods within the two windows could be ruled out, but we could highly constrain the possibilities. Future observations could be performed at the times given in this work to finally determine the definitive orbital period and confirmed the presence of the companion.

We performed a second joint fit analysis, adding the new LCO data, and obtained the final parameters distributions. The companion would have a mass of $M = 9.4^{+4.9}_{-4.1}$ $\mathrm{M_{Jup}}$ ($3\sigma$ upper limit $< 28.0$ $\mathrm{M_{Jup}}$), a radius of $R = 1.7 \pm 0.1$ $\mathrm{R_{Jup}}$, and a density of $\rho = 2.2 \pm 1.3\,\mathrm{g\,cm^{-3}}$. The mass detection result, 2.3 $\sigma$, should be treated with caution. The potential companion would lie on the edge between a giant planet and a brown dwarf, but when comparing it to the sample of transiting brown dwarfs detected so far (T. W. Carmichael 2023), the object would have a lower mass and density than them. It would be a cool object ($T_{\rm eq} = 268 \pm 13$ K), with a similar temperature to that of objects with similar periods and masses, but the only one of those that has been seen transiting. The potential companion orbits around Tabby's star in a slightly eccentric orbit, $e = 0.09^{+0.08}_{-0.05}$, at $a = 2.35 \pm 0.17$ au from it. This configuration agrees with the giant planet distribution and would explain the lack of an occultation in the data.

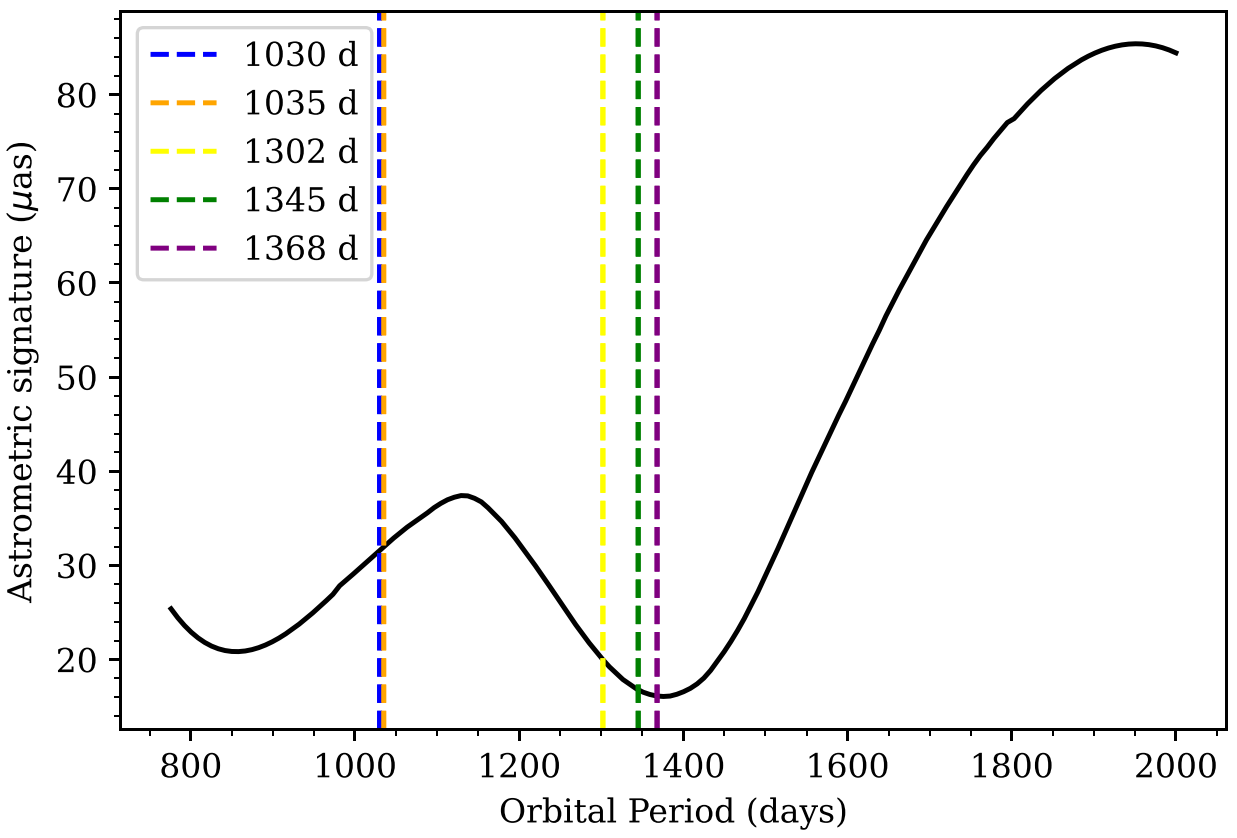

**Figure 14.** Astrometric signature of the companion as a function of the orbital period.

Other scenarios about the nature of the potential companion have been studied, although none of them seems to be supported by the observations. The scenario of it being part of an M-dwarf binary together with the nearby M-dwarf around Tabby's star does not appear to be favourable according to the *Gaia* photometry; and the transit being caused by a dust cloud or debris is highly improbable, as these are often irregular, and the transit is highly symmetrical. Exocomets models struggle to produce a good fit of the transit, which suggests that the transit is not caused by a chain of exocomets either. The presence of dust around the companion cannot be ruled out, as *TESS* data show a brightening centred on the transit, suggesting that it might be caused by forward scattering of dust around the companion. A consequence of dust around the companion would mean it would appear larger than it actually is.

In the future, upcoming *Gaia* DR4 will be able to detect the signal of a companion with the characteristics described in this work by astrometry. Its astrometric signature, the observable for astrometric planet detection, is $\alpha = 50.4 \pm 16.3$ µas, which translates to a S/N of $1.5 \pm 0.5$, a detectable signal (M. Perryman et al. 2014). Following the same procedure as in Section C, we obtained the astrometric signature we would get for different periods of the companion (Fig. 14).

*JWST* and *HST* observations could be used to observe and study the primordial atmosphere of the companion, and, along with the discovery of the next transit, they would help to disclose more information about this system.

## ACKNOWLEDGEMENTS

We thank the anonymous reviewer for their helpful comments, which have improved the paper. CMF acknowledges support from the Science and Technology Facilities Council (STFC) (research grant number ST/Y509693/1). PAS and TGW thanks funding from the University of Warwick and UKSA. This paper includes data collected by the *TESS* mission. Funding for the *TESS* mission is provided by the NASA's Science Mission Directorate. This work makes use of observations from the Las Cumbres Observatory global telescope network. This research has made use of the NASA Exoplanet Archive, which is operated by the California Institute of Technology, under contract with the National Aeronautics and Space Administration under the Exoplanet Exploration Programme.

## DATA AVAILABILITY

The *TESS* and *Kepler* data used in this work are publicly available. The rest of the data underlying this article will be shared on reasonable request to the corresponding author.

## APPENDIX A: RADIAL VELOCITY ANALYSIS

**Table A1.** Masses and periods obtained from the joint-fit using different FWHM values.

| FWHM (kms$^{-1}$) | Mass ($M_{\rm Jup}$) | Period (d) |
|---|---|---|
| Free | $11.3^{+5.8}_{-5.2}$ | $1180^{+124}_{-72}$ |
| 10 | $11.6^{+6.6}_{-6.0}$ | $1236^{+136}_{-93}$ |
| 7 | $12.4^{+5.7}_{-5.0}$ | $1211^{+89}_{-78}$ |

## APPENDIX B: OBSERVATIONS

**Table B1.** Radial velocity observations.

| Time (BJD-2457000) | RV (ms$^{-1}$) | Uncertainty RV (ms$^{-1}$) |
|---|---|---|
| 306.38574 | 18.9 | 62.5 |
| 332.39083 | 456.9 | 90.5 |
| 333.35073 | −133.5 | 86.2 |
| 732.23403 | −67.8 | 180.4 |
| 733.23429 | −77.4 | 122.2 |
| 827.67683 | −238.5 | 102.3 |
| 828.67222 | −37.0 | 98.7 |
| 829.67952 | −145.9 | 130.3 |
| 831.68694 | −382.4 | 118.9 |
| 3282.21179 | −203.3 | 124.6 |
| 3282.22670 | 10.1 | 107.1 |
| 3282.24338 | −98.9 | 100.3 |
| 3282.26017 | 75.4 | 91.4 |
| 3282.27685 | 117.6 | 92.9 |
| 3282.29347 | −74.5 | 102.1 |
| 3282.31017 | −648.9 | 112.9 |
| 3282.32800 | −190.9 | 115.5 |
| 3282.34590 | −859.0 | 167.0 |
| 3282.36259 | 302.4 | 196.8 |
| 3284.22247 | −545.4 | 140.2 |
| 3284.24058 | −119.4 | 157.3 |
| 3284.25758 | 236.4 | 155.2 |
| 3284.27498 | −301.0 | 176.7 |
| 3284.29176 | −155.5 | 205.0 |
| 3284.30844 | −203.2 | 180.1 |
| 3511.47768 | 135.3 | 88.0 |
| 3528.46467 | −237.2 | 86.3 |
| 3542.49988 | 531.0 | 98.1 |
| 3547.42601 | 23.9 | 91.0 |
| 3556.50040 | −304.3 | 102.1 |
| 3571.42520 | −138.4 | 161.3 |
| 3588.45576 | −163.4 | 97.3 |
| 3603.36599 | 44.5 | 83.3 |
| 3614.38613 | 38.6 | 100.2 |
| 3619.32302 | 488.1 | 111.3 |
| 3625.25104 | −400.6 | 90.5 |
| 3631.28655 | 262.6 | 140.4 |
| 3637.23015 | 34.0 | 178.0 |
| 3644.25257 | 128.2 | 103.1 |
| 3656.22722 | −47.8 | 97.7 |
| 3661.24844 | −36.3 | 105.0 |
| 3666.24497 | 418.8 | 149.0 |
| 3733.70673 | 445.8 | 187.1 |

**Table B2.** LCO photometry observations.

| Time (BJD-2457000) | Flux | Uncertainty Flux |
|---|---|---|
| 3789.952825 | 0.999976 | 0.0003 |
| 3790.012219 | 0.9998 | 0.0002 |
| 3790.034632 | 0.9998 | 0.0001 |
| 3790.072600 | 0.9999 | 0.0002 |
| 3790.093667 | 1.0004 | 0.0002 |
| 3790.628588 | 1.0001 | 0.0002 |
| 3790.634670 | 1.0003 | 0.0001 |
| 3790.666198 | 0.9998 | 0.0002 |
| 3790.666578 | 1.0000 | 0.0002 |
| 3790.713940 | 0.9997 | 0.0002 |
| 3790.953448 | 0.9999 | 0.0002 |
| 3791.623230 | 0.9998 | 0.0003 |
| 3791.658942 | 1.0002 | 0.0003 |
| 3791.710556 | 1.0003 | 0.0002 |
| 3792.616209 | 0.9996 | 0.0004 |
| 3792.644286 | 1.0023 | 0.0004 |
| 3792.701813 | 0.9997 | 0.0002 |
| 3798.599969 | 1.0000 | 0.0001 |
| 3798.637221 | 1.0002 | 0.0002 |
| 3798.678964 | 0.9997 | 0.0002 |
| 3798.707300 | 0.9995 | 0.0003 |
| 3799.058738 | 0.9984 | 0.0004 |
| 3799.096567 | 0.9998 | 0.0002 |
| 3799.597117 | 1.0004 | 0.0002 |
| 3799.624428 | 0.9995 | 0.0001 |
| 3799.697155 | 1.0000 | 0.0001 |
| 3799.926899 | 0.9997 | 0.0002 |
| 3799.946873 | 1.0002 | 0.0002 |
| 3799.972463 | 1.0006 | 0.0002 |
| 3800.010446 | 0.9997 | 0.0002 |
| 3800.056484 | 1.0002 | 0.0001 |
| 3800.094505 | 0.9998 | 0.0002 |
| 3800.583017 | 0.9997 | 0.0004 |
| 3800.621144 | 1.0000 | 0.0002 |
| 3800.666350 | 1.0000 | 0.0004 |
| 3800.704348 | 1.0001 | 0.0002 |
| 3800.724101 | 0.9959 | 0.0002 |

## APPENDIX C: CONSTRAINTS ON THE COMPANION MASS

T. S. Boyajian et al. (2016) obtained the upper mass limit of a potential planetary or brown dwarf companion using FIES data: the RVs and the Ellipsoidal Light Variations (ELVs) from the Fourier transform. For each in a sequence of orbital periods from 0.5 to 3000 d, they fit the RV data with a sine and cosine term assuming a circular orbit, and then used the semi-amplitude $K$ plus its $2\,\sigma$ uncertainty to obtain a conservative upper limit of the mass. Three different inclination angles were considered (30°, 60°, 90°). These results indicate that it is not likely that a companion more massive than a brown dwarf exists for periods $P \lessapprox 300$ d, but it could be present for longer periods.

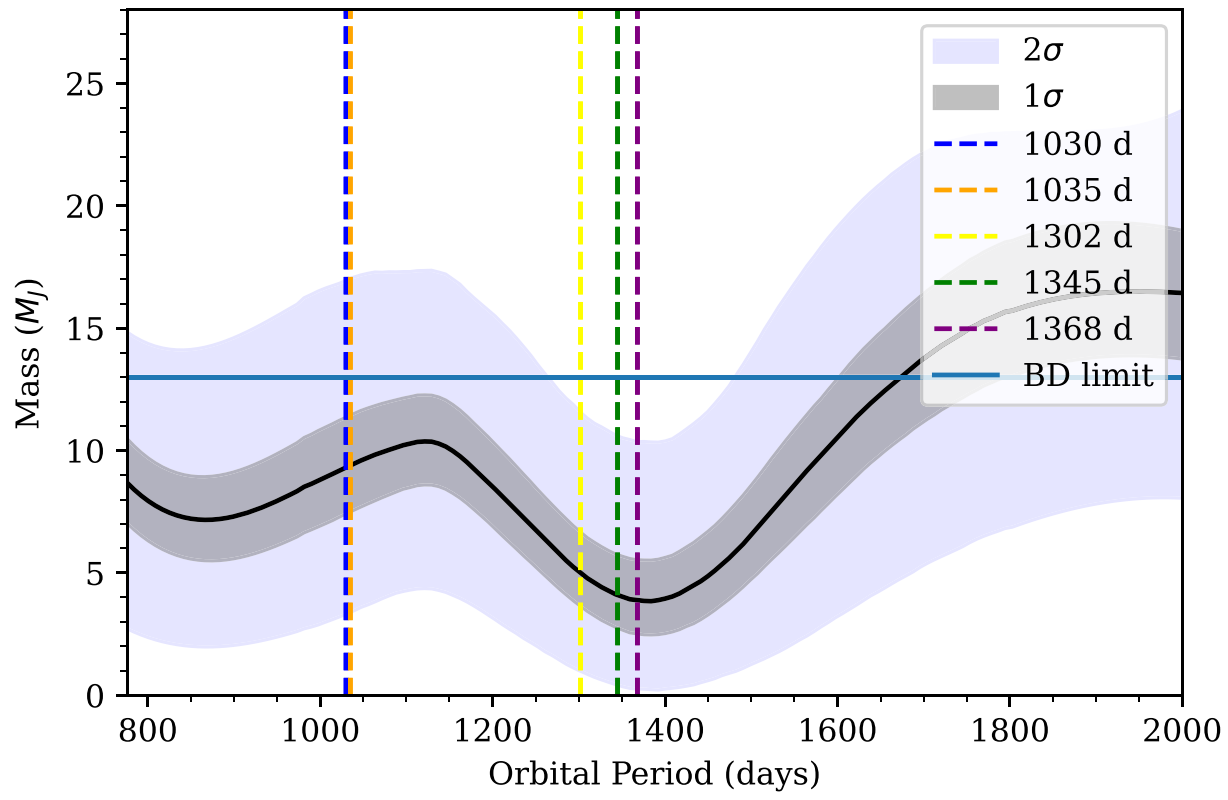

**Figure C1.** Lower and upper mass limits (at 1 and 2$\sigma$ confidence) of the companion, with an inclination of the orbits of 90°, and eccentricity of 0. The average mass is shown as the solid line.

For our analysis, we fixed the orbital period from 776 and 2000 d. For each period, we bootstrapped the SOPHIE data to produce 500 unique RV datasets and fit each to a sine function. After 500 iterations, we computed the semi-amplitude distribution, from which we obtained the mass distribution. We then took the mean, and the 1$\sigma$ and 2$\sigma$ upper and lower values of the distribution (Fig. C1). We set the bounds of the semi-amplitude to 0 and 500 m s$^{-1}$, and we assumed a circular orbit and an inclination of 90°. We must recall that, due to the fixed inclination of the orbit, these are limits of the minimum mass value of the potential companion. This method allowed us to obtain initial estimates of the mass for a big range of periods. We found that if the transit takes place during windows 1 and 2, the companion could fall into two categories: a giant planet or a brown dwarf; while for window 3, it is in the limit between the two type of objects; for windows 4 and 5, the upper mass limit sits below 13 $M_{Jup}$, and it is unlikely it is a brown dwarf. For the 2$\sigma$ lower limit, a transiting companion could have a mass as small as Jupiter.

## APPENDIX D: EXOCOMETS ANALYSIS

**Table D1.** Priors and posteriors of the parameters of the best-fitting exocomet of the *TESS* transit for Model 1. The transit duration is $t_{tr} = 0.82$ d. The index $i$ goes from 0 to number of exocomets ($n_{exocomets}-1$). The model uses three exocomets.

| Parameter | Description | Prior | Posterior |
|---|---|---|---|
| $t_{beg}$ (BJD) | Time of the beginning of the transit | $t_0 - (t_{tr}/2) + (t_{tr}/n_{exocomets}) \times i \pm 10^{-4}$ | $2458729.754 + (0.082 \times i)$ |
| $\Delta t$ (d) | Transit duration to cover the chord length of the stellar disk at the transit velocity | $(0, t_{tr}/2)$ | $0.16 \pm 0.01$ |
| | | | $0.13 \pm 0.06$ |
| | | | $0.32 \pm 0.02$ |
| $\beta$ ($d^{-1}$) | Speed of transit of one scale length of the cometary tail | (0, 0.5) | $0.36 \pm 0.07$ |
| | | | $0.43 \pm 0.14$ |
| | | | $0.39 \pm 0.07$ |
| $K$ ($10^{-4}$) | Cloud optical thickness at the leading head | $(0, 0.012 + 0.005)$ | $0.20 \pm 0.12$ |
| | | | $0.05 \pm 0.13$ |
| | | | $0.10 \pm 0.06$ |

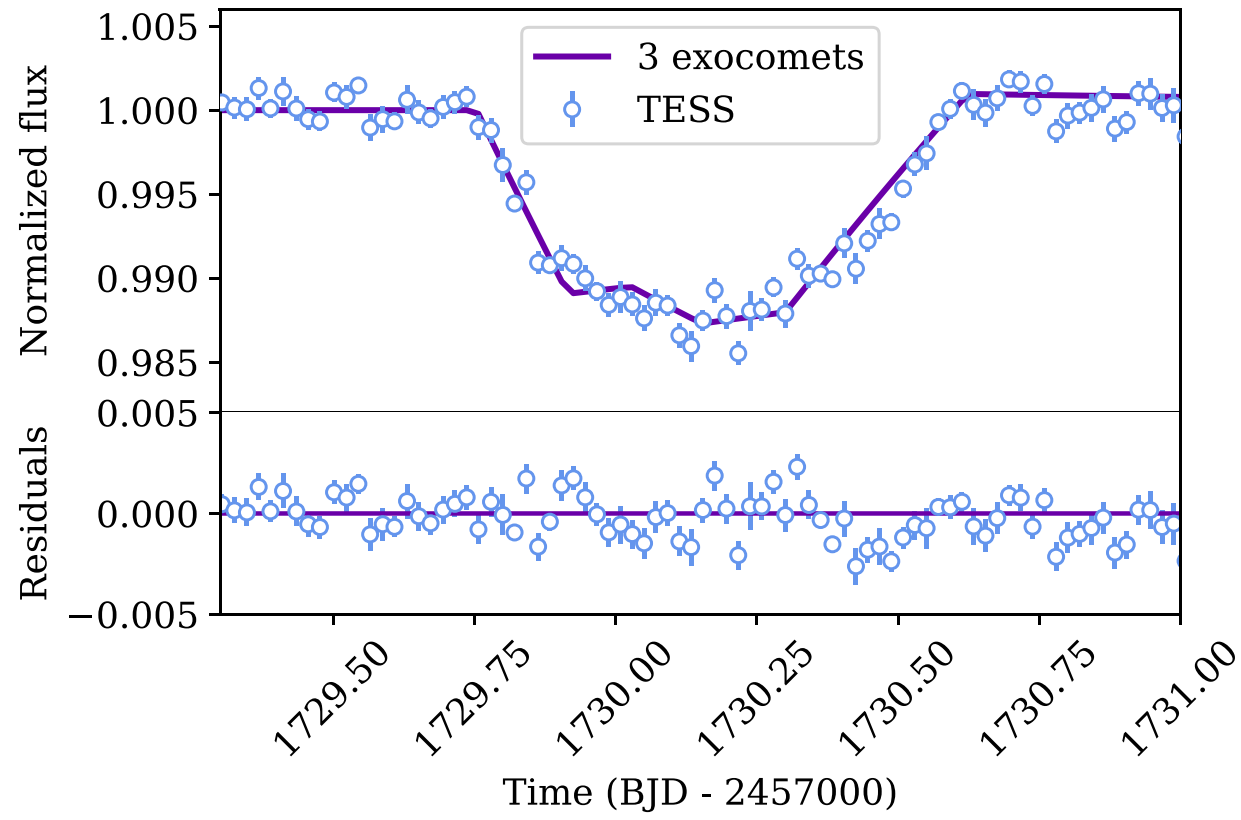

**Figure D1.** *TESS* transit with the best-fit model using Model 1, and the residuals. The model uses three exocomets.

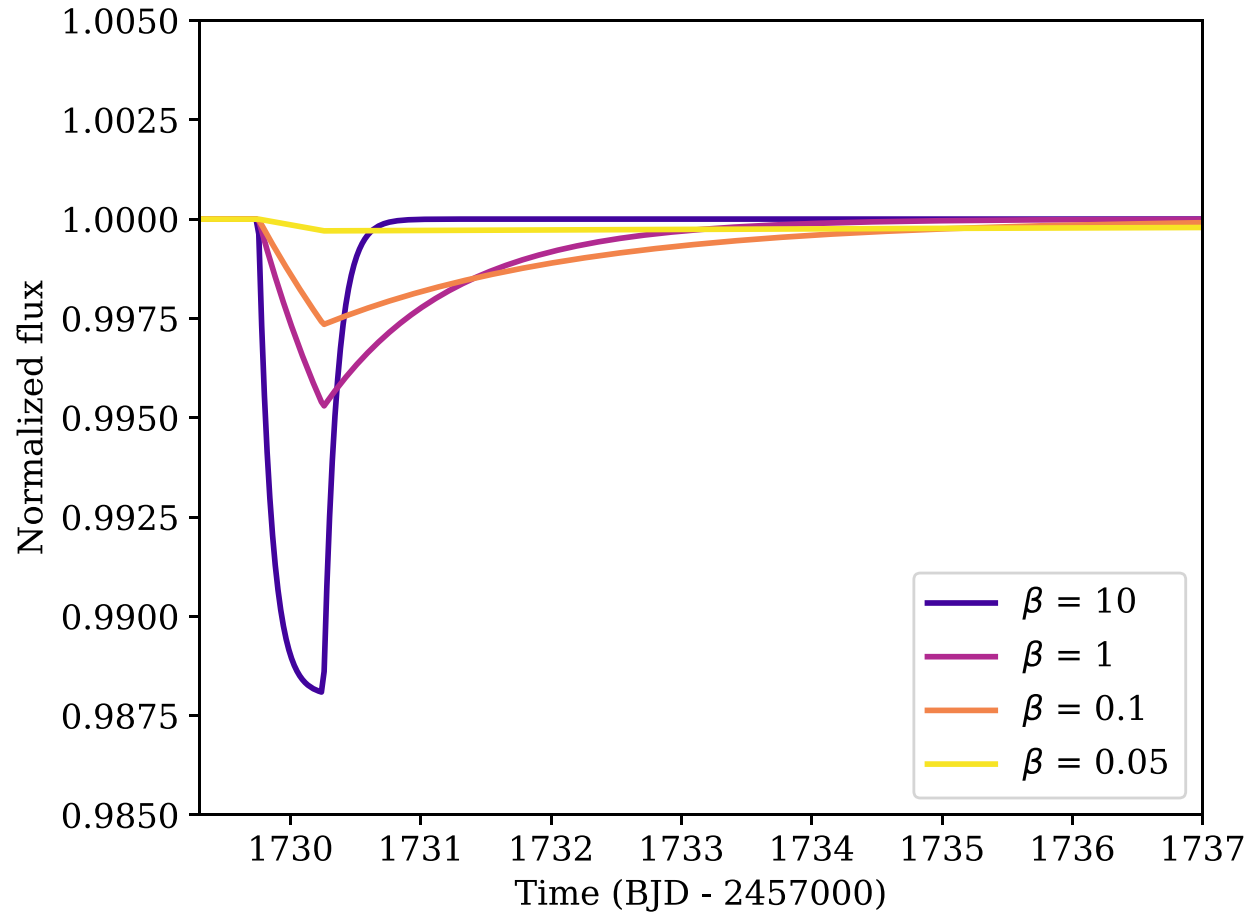

**Figure D2.** Four exocomets models (using Model 1), each with a different $\beta$ value. The smaller $\beta$ is, the flatter the brightness profile of the exocomet is.

## APPENDIX E: PHASE-FOLDED PHOTOMETRY DATA

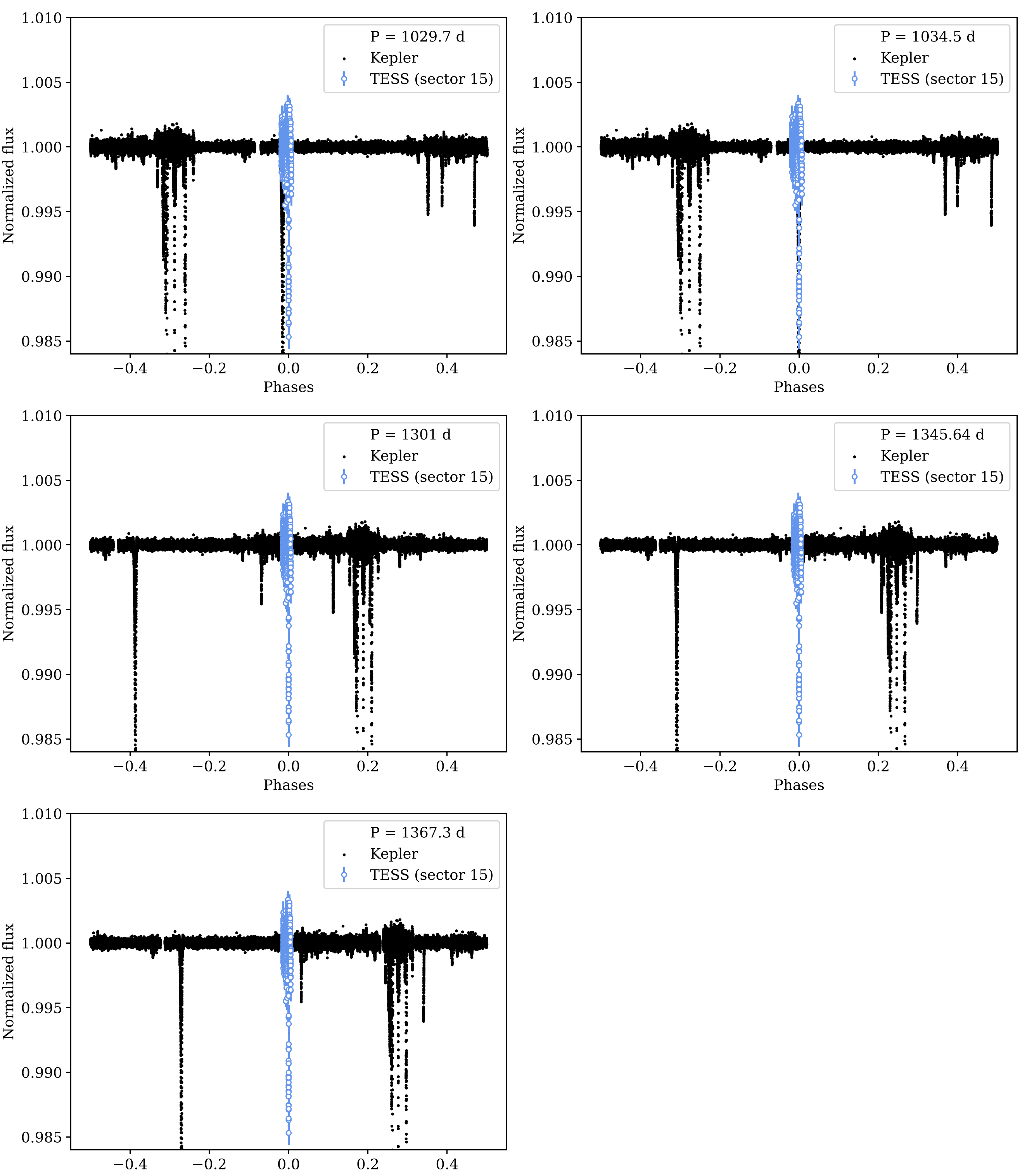

**Figure E1.** *TESS* data from sector 15 (where the transit occurred) and Kepler data phase-folded to the five potential periods. The distribution in phases of the dips in the *Kepler* data is shown, but further analysis is required to relate them to the *TESS* transit. This is beyond the scope of this paper.

This paper has been typeset from a TEX/LATEX file prepared by the author.